\documentclass[authoryear,preprint,12pt]{elsarticle}

\usepackage{url,natbib}

\usepackage{amsmath, amssymb, amsfonts, amsthm, fouriernc}

\usepackage{amsmath}
\usepackage[usenames,dvipsnames]{color}%

\newcommand{\Avec}{\mathbf{A}}
\newcommand{\Pvec}{\mathbf{P}}
\newcommand{\Vvec}{\mathbf{V}}
\newcommand{\dvec}{\mathbf{d}}
\newcommand{\evec}{\mathbf{e}}
\newcommand{\kvec}{\mathbf{k}}
\newcommand{\nvec}{\mathbf{n}}
\newcommand{\rvec}{\mathbf{r}}
\newcommand{\uvec}{\mathbf{u}}
\newcommand{\xvec}{\mathbf{x}}
\newcommand{\yvec}{\mathbf{y}}
\newcommand{\zvec}{\mathbf{z}}
\newcommand{\svec}{\mathbf{s}}
\newcommand{\Evec}{\mathbf{E}}
\newcommand{\Bvec}{\mathbf{B}}
\newcommand{\Jvec}{\mathbf{J}}

\newcommand{\del}{\partial}

\newcommand{\0}{\mathbf{0}}

\DeclareMathOperator{\diag}{{\rm diag}}

\numberwithin{equation}{section}

\journal{Foundations of Physics}

\begin{document}

\begin{frontmatter}



\title{Neo-classical Relativistic Mechanics Theory for Electrons that Exhibits Spin, Zitterbewegung, Dipole Moments, Wavefunctions and Dirac's Wave Equation}

\author[label1]{James L. Beck}
\address[label1]{California Institute of Technology, Mail Code 9-94, Pasadena. CA 91125 
\newline  E-mail: jimbeck@caltech.edu}


\begin{abstract}
%
%
In this work, a neo-classical relativistic mechanics theory is presented where the spin of an electron is an inherent part of its world space-time path as a point particle. 
The fourth-order equation of motion corresponds to the same covariant Lagrangian function in proper time as in special relativity except for an additional spin energy term.  
The theory provides a hidden-variable model of the electron where the dynamic variables give a complete description of its motion,  
giving a classical mechanics explanation of the electron's spin, its dipole moments, and Schr\"{o}dinger's zitterbewegung, 
These features are also described mathematically by quantum mechanics theory, of course, but without any physical picture of an underlying reality. 
The total motion of the electron can be decomposed into a sum of a local spin motion about a point and a global motion of this point, called here the spin center.  
The global motion is sub-luminal and described by Newton's Second Law in proper time, the time for a clock fixed at the spin center, 
while the total motion occurs at the speed of light c, consistent with the eigenvalues of Dirac's velocity operators having magnitude c. 
The local spin motion is an inherent perpetual motion, which for a free electron is periodic at the ultra-high zitterbewegung frequency 
and its path is circular in a spin-center reference frame.
In an electro-magnetic field, this spin motion generates magnetic and electric dipole energies through the Lorentz force on the electron's point charge. 
The \emph{electric} dipole energy corresponds to the spin-orbit coupling term involving the electric field that appears in the corrected 
Pauli non-relativistic Hamiltonian, which has long been used to explain the doublet structure of the spectral lines of the excited hydrogen atom. 
Pauli's spin-orbit term is usually derived, however, from his \emph{magnetic} dipole energy term, including also the effect of Thomas precession, which halves this energy. 
The magnetic dipole energy from Pauli's and Dirac's theory is twice that in the neo-classical theory, a discrepancy that has not been resolved. 
By defining a spin tensor as the angular momentum of the electron's total motion about its spin center, the fundamental equations of motion can be re-written in 
an identical form to those of the Barut-Zanghi electron theory.
This allows the equations of motion to be expressed in an equivalent form involving operators applied to a state function of proper time satisfying 
a neo-classical Dirac-Schr\"{o}dinger spinor equation. 
This state function produces the dynamic variables from the same operators as in Dirac's theory for the electron but without any probability implications. 
It leads to a neo-classical wave function that satisfies Dirac's relativistic wave equation for the free electron by applying 
the Lorentz transformation to express proper time in the state function in terms of an observer's space-time coordinates,  
showing that there is a close connection between the neo-classical theory and quantum mechanics theory for the electron's dynamics. 

\end{abstract}


\end{frontmatter}



%

\section{Introduction}


A physical explanation of the electron's spin has remained unsettled since its existence was first proposed by  {\color{Green} Uhlenbeck and Goudsmit (1925, 1926)}, 
despite the development by {\color{Green} Dirac (1928)} of a relativistic wave equation for the electron that predicted its spin and associated magnetic moment. 
It is readily shown that a simple classical model of a spherical electron spinning about an axis requires a spherical radius that is many orders larger than what is implied 
by scattering experiments. This is necessary in order to produce the correct angular momentum of $ \hbar / 2$ without surface speeds exceeding the speed of light.

In this work, we present a neo-classical relativistic mechanics model for an electron that extends Einstein's special relativity to include its spin as a natural part 
of its space-time path as a \emph{point} particle. 
The theory for this model is built primarily on earlier work of {\color{Green} Barut and Zanghi (1984), Hestenes (1985,1990, 2010), Rivas (1989, 1994, 2001, 2003, 2008)} and 
{\color{Green} Salesi (2002, 2005)}, with new connections between their work and new results, including a classical derivation of Dirac's wave equation for a free electron for the first time. 

In the model, the space-time path of an electron of mass $m$ is the sum of the motion of an auxiliary point, the \emph{spin center}, that describes the \emph{global} motion 
corresponding to the bodily transport of the electron, plus an inherent \emph{local} spin motion about the spin center. 
This perpetual spin motion is the source of the phenomenon described by {\color{Green} Schr\"{o}dinger (1930)} as \emph{zitterbewegung} 
(trembling motion) that he derived from the free electron solution of Dirac's wave equation. 
For a free electron viewed from its \emph{rest frame} fixed at the spin center so that the global velocity of the electron 
is zero, the spin motion is circular at the speed of light $c$ with angular frequency $\omega_0 = 2mc^2 / \hbar$ and radius $r_0 = c/\omega_0 = \hbar/(2mc)$, 
half of the reduced Compton wavelength, giving an angular momentum about the spin center of $mcr_0 = \hbar / 2$. 
The  \emph{rest energy} $mc^2 =  \hbar \omega_0 /2$ of the electron is the kinetic energy of this local spin motion.
Proper time $\tau$ is the time measured by a clock fixed in this rest frame. 
Note that a reference frame attached to the electron itself could never be an inertial frame because the electron is always accelerating relative to any observer inertial frame. 

In Section 2, we introduce the fundamental equations of motion for the electron for the neo-classical theory that are essentially those of {\color{Green} Rivas (2003)} 
but expressed using 4-vectors and proper time $\tau$ rather than observer time $t$.  All physical quantities are completely specified in a similar way to classical 
mechanics (hence the terminology ``neo-classical"). 
We show that the 4-vector equation for the \emph{total} (local plus global) motion of the electron is a fourth-order differential equation in proper time $\tau$ 
that is obtained by simply adding a spin energy term to the Lagrangian function from special relativity.
This equation of motion can be expressed as two coupled second-order differential equations, one for the position of the point electron and the other for its spin center. 
The latter equation describing the sub-luminal motion of the spin center is just Newton's Second Law using the Lorentz force on the point charge. 
The inertial resistance to acceleration of the electron is therefore associated only with its \emph{global} motion and there is no requirement for a centrifugal force to produce 
the inherent local spin motion. We show that a coupling between spin and the electromagnetic field arises naturally from the Lorentz force acting on the spin motion, 
leading to a neo-classical energy equation with a spin-field interaction term that is absent in Einstein's special relativity.
The spin motion also creates an apparent wave action when viewed through a Lorentz transformation from an observer inertial reference frame. 

In Section 3, we introduce a spin tensor $S$ representing the angular momentum from the total velocity about the spin center. 
Using this tensor, the equations of motion can be re-written in a form that mimics the operator equations derived from the Heisenberg picture of the proper-time Dirac wave 
equation given in {\color{Green} Barut (1987)}. The same equations have been derived by others using several different approaches but not by starting with the fundamental 
neo-classical equations of motion presented in  Section 2 as done here (e.g. {\color{Green} Weyssenhoff and Raabe (1947), Grossmann and Peres (1963), 
Barut and Zanghi (1984), Salesi (2002)}, and {\color{Green} Hestenes (2010)}). 
Using the spin tensor, the spin-field interaction term in the neo-classical energy equation can be expressed as the sum of magnetic and electric dipole energies 
that are exactly a half of Dirac's theory. The electric dipole term is consistent, however, with the spin-orbit coupling exhibited by 
the corrected Pauli non-relativistic Hamiltonian that explains the doublet structure of the spectral lines of the excited hydrogen atom without calling on Thomas precession.

In Section 4, a neo-classical four-dimensional complex \emph{state function} is introduced that, when used in the usual Hermitian inner products 
with self-adjoint linear operators, produces the \emph{actual} dynamic quantities (not their expected values) as a function of proper time. 
The evolution of this state function is described by a first-order differential equation with respect to proper time given by the spinor equation in the theory of 
{\color{Green} Barut and Zhangi (1984)}. This equation is similar to the Feynman proper-time parameterization of the Dirac-Schr\"{o}dinger equation in QM (quantum mechanics). 
We show that the spin tensor form of the equations of motion can be expressed in operator form by using the Dirac velocity and spin tensor operators that act on the spinor state function. 
In doing so, we show consistency between the neo-classical spin tensor $S$ and the QM spin tensor operator. 

Also in Section 4, we show that in the rest frame for a free electron, a superposition of the spin-up and spin-down neo-classical state functions produces the state function 
for spin about an axis whose direction is specified by the superposition coefficients. 
The implied velocities from the Dirac velocity operators (the Dirac matrices) are shown to lie in a plane perpendicular to the spin axis and agree with the basic hypothesis 
of the neo-classical theory that the electron's spin motion is circular at the speed of light $c$ with  angular frequency $\omega_0 = 2mc^2 /\hbar$. 
These spin state functions are in turn superpositions of the so-called ``positive energy'' and ``negative energy'' solutions, which actually both have positive energy; 
the latter are just the conjugate solutions needed to properly produce the periodic spin motion under the action of the velocity operators.

Finally in Section 4, we show by using a Lorentz transformation for proper time that the state function solutions of the neo-classical Dirac- \linebreak Schr\"{o}dinger equation 
for a free electron become neo-classical \emph{wave functions} that satisfy Dirac's wave equation where the independent variables are the 4-vector position coordinates 
with respect to an observer inertial reference frame. 
Thus, for the first time, Dirac's wave equation is derived directly from a modified classical relativistic mechanics theory that has spin incorporated. 
These wave functions exhibit wave-like behavior with a frequency and wavenumber proportional to the electron's energy and global kinetic momentum as in de Broglie's theory.
However, minimal coupling in the neo-classical Lagrangian function does not produce the same field coupling implied by the usual QM minimal coupling prescription in Dirac's 
equation since it produces one half of the magnetic dipole energy for the electron compared with Dirac's derivation. 
Thus, the neo-classical theory provides a consistent hidden-variable model for spin and Dirac's relativistic wave equation for the free electron but it needs further development 
to improve the coupling between an electron and an electro-magnetic field. 

In Section 5, some remarks are made about introducing probability into the neo-classical theory since its wave functions currently have no probability implications, 
unlike in QM theory. Following concluding remarks in Section 6, three appendices give some details of a few results stated without proof in the main sections.

\section{Neo-Classical Spin Model and its Fundamental Equations}

In this section, the fundamental equations of motion for an electron in an em-field (electromagnetic field) that describe the neo-classical model for spin are presented, 
along with an explicit solution of these equations for the free electron showing that its dynamics naturally exhibits spin through its space-time motion and that 
Schr\"{o}dinger's ZBW (zitterbewegung) is just a manifestation of this spin motion.
We derive two important energy equations and then show that the equations of motion can also be derived by adding a term for the spin kinetic energy to the usual classical 
relativistic Lagrangian function for the electron.

\subsection{Equations of Motion}

Using 4-vectors relative to an observer inertial reference frame $X_o$ with its origin at $O$, the \emph{total} motion $x(\tau)$ of the electron of mass $m$ and charge $q = -e$ 
(so $e>0$ is the unit electronic charge) is modeled as the sum of: 
(i) a \emph{local} spin motion $z(\tau)$ where the electron moves in perpetual motion about its spin center $C$, and 
(ii) a \emph{global} motion $y(\tau)$ of the electron corresponding to the motion of this spin center. 
In the \emph{rest frame} for the global motion, which corresponds to a reference frame $X_c$ fixed at $C$ (not inertial if there is an electromagnetic field present), the 
trajectory of a free electron is a circular path traversed at the speed of light $c$ with the ZBW angular frequency $\omega_0 = 2mc^2 / \hbar \sim 1.55 \times 10^{21}$s$^{-1}$ 
and its kinetic energy is $mc^2 = \frac{1}{2} \hbar \omega_0$, the rest energy of Einstein's special relativity. 
The radius of this circular spin motion is $r_0 = c/ \omega_0 = \hbar/(2mc) \sim 1.93 \times 10^{-13}$m so that the spin angular 
momentum of the electron is $s = r_0mc = \hbar/2 = h/(4\pi)$ where $h$ denotes Planck's constant. 

No centrifugal force is required for the spin motion, which is an inherent perpetual motion, implying that the electron's mass, physically located at its position $x$, 
is not directly an inertial mass. Instead, the inertia to moving the electron comes from accelerating the  \emph{global} motion, $y(\tau)$, of the spin center $C$ 
(also called the \emph{center of mass} for this reason), which is described by Newton's Second Law using \emph{proper time} $\tau$, the time for a clock fixed at $C$. 
Clearly, an electron must only emit and absorb electromagnetic energy when the global motion is accelerating. 
Furthermore, the spin motion, $z(\tau)$, cannot be observed directly because it is much too fast and so the location of the electron cannot be known more accurately 
than a position vector error of magnitude $r_0 = \hbar/(2mc)$. 
The spin motion $z(\tau)$ is essentially a hidden variable but it has consequences that are observable, such as the electron's magnetic and electric dipole moments. 

The equations of motion for the four components of the total and global motions, $x(\tau)$ and $y(\tau)$, are therefore:
\begin{align}\nonumber
	\ddot{x}^{\mu} & = - \omega_0^2 \left( x^{\mu} - y^{\mu} \right) \  \\ \label{Eq_1_1}
	\ddot{y}^{\mu} & = \frac{1}{m} f^{\mu} = \frac{q}{m} F^{\mu \nu} (x) \dot{x}_{\nu}
\end{align}
\noindent 	Here $F(x)$ is the em-field tensor at the electron's space-time coordinates $x$, an overhead dot denotes a derivative with respect to proper 
time $\tau$, and $x^{\mu} (\tau)$ and $y^{\mu} (\tau)$ ($\mu = 0,1,2,3$) are the contravariant components of the 4-vector position coordinates $x = (ct,\xvec)$ for the total motion 
and $y = (ct_y,\yvec)$ for the spin center motion relative to the inertial reference frame $X_o$ at $O$. 
The local spin motion $z = (ct_z,\zvec)$ about the spin center $C$ relative to $X_o$ then has the contravariant components $z^{\mu} = x^{\mu} - y^{\mu}$. 
Throughout the paper, bold letters are 3-vectors with ${\bf a} \cdot {\bf b}$ and ${\bf a} \times {\bf b}$ denoting the scalar and vector products (${\bf a}^2 = {\bf a} \cdot {\bf a}$ ) and 
Einstein's summation convention for repeated indices is used. The Minkowski metric tensor is $G = \diag(1,-1,-1,-1)$ with components $g_{\mu\nu} = g^{\mu\nu}$, and 
the space-time inner product of $y$ and $z$ is denoted $y*z = y_{\mu} g^{\mu\nu} z_{\nu} = y_{\mu} z^{\mu} = y^0z^0 - y^1z^1 - y^2z^2 - y^3z^3$. 

Using the usual expression for the em-field tensor $F$, the Lorentz force $f=(f^0,{\bf f})$ in Eq.~\eqref{Eq_1_1} can be expressed as:
\begin{equation} \label{Eq_1_1b}
	f^0 = \frac{q}{c} \Evec \cdot {\bf \dot{x}} \text{  and  } {\bf f} = q \left( \dot{t} \Evec + {\bf \dot{x}} \times  \Bvec \right)
\end{equation}
\noindent where $ \Evec (x)$ and $ \Bvec (x)$ are the electric and magnetic field strengths and SI units are used. 
Introducing the 4-vector \emph{global} momentum $\pi = m \dot{y} = \left( \frac{E}{c}, \Pvec \right)$, the last equation in Eq.~\eqref{Eq_1_1} can be written as $\dot{\pi} = f$ and so:
\begin{equation} \label{Eq_1_1c}
	\dot{E} = c f^0 = q \Evec \cdot {\bf \dot{x}} \text{  and  } \dot{\Pvec} = {\bf f} = q \left( \dot{t} \Evec + \dot{\xvec} \times  \Bvec \right)
\end{equation}
\noindent implying that $\dot{E} = \dot{\Pvec} \cdot \dot{\xvec}/ \dot{t} = {\bf f}  \cdot \frac{d \xvec}{dt}$, 
so the proper time rate of kinetic energy change is equal to the observed rate of work done by the em-field on the electron, as in special relativity.

The equations of motion in Eq.~\eqref{Eq_1_1}, which are remarkably simple when expressed using proper time $\tau$, require specification of 16 initial conditions: 
$x(0)$, $\dot{x}(0)$, $y(0)$, $\dot{y}(0)$, for a unique solution. In addition, there are two constraints: 
\begin{align}\nonumber
	C1: \dot{x}*\dot{x} & = \dot{x}_{\mu} \dot{x}^{\mu} = 0 \  \\ \label{C1_C2}
	C2: \ddot{x}*\ddot{x} & = \ddot{x}_{\mu} \ddot{x}^{\mu} = - c^2 \omega_0^2
\end{align}
\noindent 	Constraint $C1$ specifies that the electron always moves at the speed of light $c$. 
Constraint $C2$ implies that a free electron travels in a circle of radius $r_0 = c/ \omega_0$ relative to the rest frame at $C$ with 
angular frequency $\omega_0$. Under Eq.~\eqref{Eq_1_1}, $C2$ can be expressed as $z*z = - r_0^2$. 
It is hypothesized that the motion of all fundamental particles with spin satisfy these two constraints with an appropriate choice of $\omega_0$. 

The equations of motion have several interesting features. 
The first equation expresses the kinematics for the electron's acceleration in terms of what the inertial observer fixed at O sees of the local spin motion $z(\tau)$. 
It can be re-written as: $\ddot{z}^{\mu} + \omega_0^2 z^{\mu} = - \ddot{y}^{\mu}$ and represents the phenomenon of ZBW revealed by {\color{Green}Schr\"{o}dinger (1930)} 
in his analysis of Dirac's wave equation. He showed that for a free electron, the ``positive'' and ``negative'' energy solutions of the wave function ``interfere'' to produce 
a motion with the high frequency $\omega_0 = 2 mc^2 / \hbar$ (twice de Broglie's frequency) and with a spatial extent of $r_0 = c/\omega_0 = \hbar/ (2mc)$ 
related to Compton's wavelength. We will show that for a free electron, the equations in Eq.~\eqref{Eq_1_1} ultimately imply Dirac's wave equation without any explicit 
quantization, giving it a derivation from a classical relativistic mechanics theory for the first time. 

In the second equation of Eq.~\eqref{Eq_1_1}, Newton's Second Law describes the sub-luminal motion of the spin center that gives the \emph{global} motion of the electron 
where the Lorentz force $f$ due to the em-field acts directly on the charge and so it is evaluated at the position $x$ of the electron. 
The \emph{total} (local plus global) motion $x(\tau)$ of the electron is described by an equivalent single fourth-order differential equation given later, 
which implies that the local spin motion is inherent and so there is no requirement for a centrifugal force to produce it. 
The equations reveal that the inertia is not a property of the mass per se but instead it is associated with resistance to accelerating the ``ring'' of the electron's spin motion. 
The spin center $C$ therefore acts like a center of mass, the terminology used by others (e.g. {\color{Green}Rivas (2001)} and {\color{Green}Salesi (2002)}).

In his book, {\color{Green}Rivas (2001)} derives the equations of motion in Eq.~\eqref{Eq_1_1} from a Lagrangian function and extensively analyzes them for the 
non-relativistic case where the overhead dot corresponds to a derivative with respect to observer time $t$, not our proper time $\tau$. 
Actually, in this non-relativistic case, the Galileo transformation replaces the Lorentz one so the time parts ($\mu = 0$) of Eq.~\eqref{Eq_1_1} are not needed because then 
$t = t_y = \tau$ in the theory presented in this paper. 
Rivas shows that the equations of motion exhibit the electron's spin, ZBW and a magnetic moment that is one half of Dirac's value, as well as the ability to jump a 
potential barrier that a classical particle without spin could not do. 
He also shows (e.g. {\color{Green}Rivas (2003)}{) that it is possible to have a bound state between 
two electrons with parallel spin directions if their circular spin trajectories partially overlap. This interesting possibility may give an explanation 
for the Cooper electron pairs postulated in a theory for ``unconventional'' (``high temperature'') superconductors.

{\color{Green}Rivas (2001)} also studies the case of an electron in a uniform magnetic field $\Bvec$ and finds that the orbital motion exhibits a circular frequency that is 
essentially the expected cyclotron frequency $\omega_c = eB/m$ but that the precession of the spin has the Larmor frequency $\frac{1}{2} \omega_c$. 
His equations for the motion are the same as the spatial part of Eq.~\eqref{Eq_1_1} except that $t$ replaces $\tau$, so his results are applicable to the relativistic spin model presented here. 
In QM theory, the spin precession frequency is $ \omega_c$ (see, for example, {\color{Green}Barut and Bracken (1982)} and {\color{Green}Barut and Thacker (1985)} who 
solve the Heisenberg equations corresponding to Dirac's equation using observer time $t$ and proper time $\tau$, respectively). 
In the non-relativistic case of {\color{Green}Rivas (2001)}, as in the relativistic spin model here, the spin-field coupling with its magnetic and electric dipole energies come 
from the Lorentz force acting on the electron's spin motion. 
We will see later that a rigorous derivation of the electron's dipole energies from a new relativistic energy equation produces one half of the magnetic dipole energy from 
Dirac's theory but appears to give an acceptable electric dipole energy expression.

For the relativistic case, {\color{Green}Rivas (2001)} investigates various Lagrangian functions, all having a similar structure but he again uses observer time $t$ rather than proper time $\tau$ used in this paper. 
One special case that is similar to the relativistic spin equations presented in this paper corresponds to the electron moving at the speed of light (called the lumen case) where 
the equations of motion derived from the Lagrangian function exhibit similar behavior to his non-relativistic case.


\subsection{Free Electron Motion}

In the case of a free electron, the equations in Eq.~\eqref{Eq_1_1} become $\ddot{z} = - \omega_0^2 z$ and $\ddot{y} = 0$. Therefore, $\dot{y}$ = constant and the 4-vector spin motion $z$ satisfies:
\begin{align} \nonumber
	z(\tau)  & = z(0) \cos (\omega_0 \tau) + \left( \dot{z}(0) / \omega_0 \right) \sin (\omega_0 \tau) \\ \label{Eq_1_3}
	\dot{z}(\tau)  & = - \left( z(0) \omega_0 \right) \sin (\omega_0 \tau) + \dot{z}(0) \cos (\omega_0 \tau) 	
\end{align}
\noindent and so $z(\tau)$ is periodic with the ZBW period $2\pi/ \omega_0$. 	

The 4-vector global momentum $\pi = m \dot{y} = \left( \frac{E}{c}, \Pvec \right)$ is constant in this free electron case and so $y(\tau) = (\pi/m) \tau + y(0)$. 
Using Eqs.~\eqref{Eq_1_1} and~\eqref{Eq_1_3}, the total 4-vector velocity and 4-vector position can be written as:
\begin{align} \nonumber
	\dot{x}(\tau) & = \dot{y}(\tau) + \dot{z}(\tau)  = \frac{1}{m} \pi + \left[ \dot{x}(0) - \frac{1}{m} \pi \right] \cos (\omega_0 \tau) + \frac{1}{\omega_0} \ddot{x}(0)  \sin (\omega_0 \tau)  \\ \label{Eq_1_4}
	x(\tau) & = y(0) + \frac{1}{m} \pi \tau + \frac{1}{\omega_0} \left[ \dot{x}(0) - \frac{1}{m} \pi \right] \sin (\omega_0 \tau) + \left[ x(0) - y(0) \right] \cos (\omega_0 \tau)
\end{align}
\noindent These solutions show that the electron's free motion in space is a helix where the circling at the speed of light and angular frequency $\omega_0$ corresponds to 
Schr\"{o}dinger's ZBW and represents the local spin contribution $z(\tau)$ while the global velocity $\dot{y} = \frac{1}{m}\pi$ is given in terms of the electron's global momentum. 
Since $x^0(\tau) = ct$, the last equation in Eq.~\eqref{Eq_1_4} shows that $t$ is a non-linear function of proper time $\tau$ along the electron's space-time path. 

The electron's kinetic energy $E = c \pi_0 = m c \dot{y}_0 = mc^2  \dot{t}_y = mc^2 \gamma$, defining $\gamma = \dot{t}_y$. 
The global momentum components for $\mu \neq 0$ are $P_{\mu} =  \pi_{\mu} = m \dot{y}_{\mu} = m \frac{dy_{\mu}}{dt_y} \frac{dt_y}{d\tau} = \gamma m V_{\mu}$ where $V_{\mu} = \frac{dy_{\mu}}{dt_y}$, 
which becomes in 3-vector notation, $\Pvec =  \gamma m \Vvec$. These results also show that $\dot{y} = \left(\gamma c, \gamma \Vvec \right)$. 
Note that $\Vvec$ is the velocity of the electron's global motion (the motion of its spin center) relative to an observer fixed with respect to the inertial reference frame $X_o$. 
For a free electron where $\pi$ is constant, $\gamma$ and $\Vvec$ are constant.
The factor $\gamma m$ is sometimes viewed as an effective relativistic mass, although $\gamma = \dot{t}_y$ is due to the different time scales, $t_y$ and $\tau$, at the center of spin $C$ when relative to 
observers fixed in the inertial reference frames $X_o$ and $X_c$, respectively.

The inner product $y*z$ is invariant under a Lorentz transformation between inertial reference frames and so it is independent of the choice of inertial reference frame used to express the 4-vectors. 
Consider the inertial reference frame $X_c$ with its origin at the spin center $C$, then $x'=(c\tau,\rvec)$ with $y'=(c\tau,\0)$ and $z'=(0,\rvec)$ the time-like and space-like components of $x'$, respectively. 
Thus, $\dot{y}*\dot{y} = \dot{y'}*\dot{y'} = (c,\0)*(c,\0) = c^2$. Substituting for $\dot{y}$ gives $\gamma = \left[ 1 - \Vvec^2/c^2 \right]^{-1/2}$, 
which corresponds to the well-known Lorentz transformation factor between the inertial reference frames $X_o$ and $X_c$. 
This is not always the case, however, because $\gamma$, which denotes $\dot{t}_y$ in this paper, gets modified by a factor when an em-field is present, as shown in the next subsection. 

These results also give the usual relativistic energy equation $m^2c^2 = m\dot{y}*m\dot{y} = \pi*\pi = E^2/c^2 - \Pvec^2$ and an additional one that is a feature of the spin model,  
$\pi*\dot{x} = m \left( \dot{y}*\dot{y} + \dot{y}*\dot{z} \right) = m \dot{y}*\dot{y} = mc^2$ where we have used the invariance of the inner product $\dot{y}*\dot{z} = \dot{y'}*\dot{z'} = (c,\0)*(0,\dot{\rvec}) = 0$.
As shown in the next subsection, the second energy equation also holds in an em-field but the first energy equation requires another term due to the coupling between the spin and the em-field. 

Using the inner product invariance between reference frames $X_o$ and $X_c$ again to evaluate the motion constraints in Eq.~\eqref{C1_C2}, we get: 
\begin{align}\nonumber
	C1: & ~0 = \dot{x}*\dot{x} = \dot{x}' *\dot{x}' = (c,\dot{\rvec})*(c,\dot{\rvec}) = c^2 - \dot{\rvec}^2 \  \\ \label{C1_C2b}
	C2: & - c^2 \omega_0^2 = \ddot{x}*\ddot{x} = \omega_0^4 z*z =  \omega_0^4 z'*z' = \omega_0^4 (0,\rvec)*(0,\rvec) = - \omega_0^4 \rvec^2 
\end{align}
\noindent  We see that the two constraints therefore imply $\dot{\rvec}^2 = c^2$ and $\rvec^2 =  r_0^2$, 
giving the consistent result that the circular motion of the free electron about the spin center $C$, relative to the rest frame $X_c$, has speed $c$ and radius $r_0 = c/ \omega_0 = \hbar/(2mc)$, respectively. 
If this circular motion lies in the $r^1-r^2$ plane, then the spatial part of one solution in Eq.~\eqref{Eq_1_3} is 

$\rvec(\tau) = \left( r_0 \cos (\omega_0 \tau + \varphi_0), r_0 \sin (\omega_0 \tau + \varphi_0), 0 \right)$ 

$\dot{\rvec}(\tau) = \left( -c \sin (\omega_0 \tau + \varphi_0), c \cos (\omega_0 \tau + \varphi_0), 0 \right)$ 

\noindent where $\varphi_0$ is the initial phase angle specifying the electron's position on the circle $\rvec^2 = r_0^2$ at $\tau = 0$. 
The spin vector due to this circular motion is $\svec = \rvec \times m \dot{\rvec} = \left( 0, 0, r_0 mc \right) = \left( 0, 0, \hbar/2 \right)$, which has magnitude $\hbar/2$ and its direction is orthogonal to the plane of circular motion. 
Therefore, this corresponds to the electron being in a spin-up state in the $x^3$ direction. Similarly, 

$\rvec(\tau) = \left( r_0 \cos (\omega_0 \tau - \varphi_0), -r_0 \sin (\omega_0 \tau - \varphi_0), 0 \right)$ 

$\dot{\rvec}(\tau) = \left( -c \sin (\omega_0 \tau - \varphi_0), -c \cos (\omega_0 \tau - \varphi_0), 0 \right)$ 

\noindent is another solution with spin vector $\svec = \rvec \times m \dot{\rvec} = \left( 0, 0, - r_0 mc \right) = \left( 0, 0, - \hbar/2 \right)$, which corresponds to the electron being in a spin-down state. 


The Lorentz transformation between the reference frames $X_o$ and $X_c$ gives alternative but equivalent expressions to those in Eqs.~\eqref{Eq_1_3} and~\eqref{Eq_1_4} for 
the time and space components (for convenience, assume that the direction of the spatial axes of $X_o$ and $X_c$ are aligned, so no rotation matrix is needed):
\begin{align} \nonumber
	 t(\tau) & =  \left[ \frac{\gamma}{c^2} \left(\Vvec \cdot \rvec(\tau) \right) \right] + \left[ \gamma \tau + a \right] = t_z(\tau) + t_y(\tau) =  \frac{1}{c} z^0(\tau) + \frac{1}{c} y^0(\tau)
	 \\ \label{Eq_1_4b}
	\xvec(\tau) & = \left[ \rvec(\tau) + \frac{\gamma^2}{(1+\gamma)c^2} \left(\Vvec \cdot \rvec(\tau) \right) \Vvec \right] + \left[ \gamma \tau \Vvec + \mathbf{b} \right] = \zvec(\tau) + \yvec(\tau) 
\end{align}
\noindent where $\ddot{\rvec} = - \omega_0^2 \rvec$ so that $\rvec(\tau) = \rvec(0) \cos (\omega_0 \tau) + \left( \dot{\rvec}(0) / \omega_0 \right) \sin (\omega_0 \tau)$, 
while $a = t_y(0)$ and $\mathbf{b} = \yvec(0)$ are constants. 
This implies that $\dot{t}(\tau) = \gamma \left( 1 + \frac{1}{c^2} \Vvec \cdot \dot{\rvec}(\tau) \right) > 0$ since $ | \Vvec \cdot \dot{\rvec}(\tau) | <c^2$, 
so $t$ always increases monotonically with $\tau$, despite its harmonic variation $t_z(\tau)$ with frequency $\omega_0$ about the linear trend $ t_y(\tau) = \gamma \tau + a $. 

Consider the inverse Lorentz transformation relating the electron's coordinates $(c \tau, \rvec)$ using the reference frame $X_c$ to the coordinates $(ct, \xvec)$ using the reference frame $X_o$:
\begin{equation}\label{Eq_1_5}
	d\tau = \gamma dt - \frac{\gamma}{c^2} \Vvec \cdot d\xvec  
\end{equation}
\begin{equation}
	d\rvec = d\xvec - \gamma \Vvec dt + \frac{\gamma^2}{c^2(1+ \gamma)} (\Vvec \cdot\ d\xvec) \Vvec
\end{equation}
\noindent The first equation can be rewritten using the previous expressions derived for the energy and 3-vector momentum, $E = \gamma mc^2$ and $\Pvec =  \gamma m \Vvec$, as:
\begin{equation}\label{Eq_1_6}
	mc^2d\tau = Edt - \Pvec\ \cdot d\xvec = \pi*dx 
\end{equation}
\noindent This equation implies the previously derived energy equation:
\begin{equation}\label{Eq_1_7}
	C3: E\dot{t} - \Pvec \cdot \dot{\xvec} = \pi*\dot{x} = mc^2 
\end{equation}
\noindent We could consider this equation to be a constraint; however, only one of the constraints $C2$ and $C3$ are independent, as shown in the next subsection. 

Since $\pi$ is constant for the free electron case, the Lorentz transformation in Eq.~\eqref{Eq_1_6} can be integrated to give 
$mc^2 \tau = \pi*x = Et - \Pvec \cdot \xvec$ (take $\tau = 0$ when $t = 0$ at $\xvec=\0$), or, equivalently, 
\begin{equation}\label{Eq_1_10b}
	\omega_0 \tau = \left(\pi*x \right)/h^* = \left(Et - \Pvec \cdot \xvec \right)/h^* = \omega t - \kvec \cdot \xvec = \gamma \omega_0 \left( t - \frac{1}{c^2} \Vvec \cdot \xvec \right)
\end{equation}
\noindent where $h^* = \hbar/2$, $\omega = E/h^* = \gamma \omega_0$ and $\kvec = \Pvec/h^* = \gamma \frac{\omega_0}{c^2} \Vvec$, 
giving the well-known de Broglie frequency-energy and wavenumber-momentum relations except for a factor of two. 
Thus, the inherent periodic spin motion with phase $\omega_0 \tau$ in Eqs.~\eqref{Eq_1_3} and~\eqref{Eq_1_4} gives an apparent plane-wave characteristic to 
the electron's motion with de Broglie's apparent wave propagation speed of $c^2/| \Vvec | > c$ when viewed by an observer fixed with respect to the reference frame $X_o$.


\subsection{Energy Equations}

Consider now the general case of an electron in an em-field. Under $C1$ and Eq.~\eqref{Eq_1_1}, the energy constraint $C3$ holds if and only if $C2$ holds, which can be shown as follows. 
Let $u=\dot{x}$ and differentiate $C1$, then $0 = u*\dot{u} = - \omega_0^2 u*z$. Then using the derivative of $u*z = 0$:
\begin{equation}\label{Eq_1_7b}
	- \frac{1}{\omega_0^2} \dot{u}*\dot{u}  = \dot{u}*z = -u*\dot{z} = -u*(u - \dot{y}) = u*(\pi/m)
\end{equation}
\noindent by using $C1$ again. Therefore, $\pi*u = mc^2$ if and only if $\dot{u}*\dot{u} = - c^2 \omega_0^2$, proving that $C2$ and $C3$ are equivalent constraints. 
We take $C1$ and $C2$ as the fundamental constraints, so $C3$ is a derived energy equation under Eq.~\eqref{Eq_1_1}, the fundamental equations of motion of the neo-classical spin model.


From $C3$: $0 = \dot{\pi}*\dot{x} + \pi*\ddot{x} = -\omega_0^2\pi*z$ by using Eq.~\eqref{Eq_1_1} and the anti-symmetry of the tensor 
$F^{\mu\nu}$ so that $ \dot{\pi}*\dot{x} = m\ddot{y}*\dot{x}  = q F^{\mu\nu} \dot{x}_{\mu} \dot{x}_{\nu} = 0$. Thus,
\begin{equation}\label{Eq_1_8}
	\pi*z = 0
\end{equation}
\noindent Differentiating: $0 = f*z + \pi*\dot{z}$. 
Using C3 again, $\pi*\dot{z} = \pi*\dot{x} - \pi*\dot{y} = mc^2 - \frac{1}{m} \pi*\pi$, giving the \emph{energy equation} that generalizes the previous free electron case:
\begin{equation}\label{Eq_1_9}
	\frac{1}{m} \left( E^2/c^2 - \Pvec^2 \right) = \frac{1}{m} \pi*\pi = mc^2 + \Phi
\end{equation}
\noindent where $\Phi = - \pi*\dot{z} = f*z$ is a spin-field interaction energy. As shown later, this term comes from the electric and magnetic dipoles set up by the spin motion of the electron. 
Eq.~\eqref{Eq_1_9} shows that the kinetic energy $E = c \pi_0$ comes from the ``rest" kinetic energy $mc^2$ and the energy $\Phi$ from the action of the Lorentz force, 
both energies coming from the spin motion, plus the kinetic energy $\frac{1}{m} \Pvec \cdot \Pvec$ from the global motion.

\subsubsection{Flywheel analogy for kinetic energy storage}

The kinetic energy of a free electron can be viewed as being stored in the spin kinetic energy as if it were a flywheel. 
For a larger global velocity of the electron relative to one inertial observer compared with another, the spin of the electron 
about its center C will appear to be faster. 
The kinetic energy associated with this spin is equal to the sum of the rest energy $mc^2$ and the translational kinetic energy due to the global velocity of the electron. 
Quantitatively, we have from the previous subsection that the apparent spin frequency of the electron's local motion for an observer fixed relative to the inertial reference frame $X_o$ is given by 
$\omega = \gamma \omega_0$ where $\omega_0$ is the spin frequency for an observer fixed relative to the electron's spin center (i.e. relative to the inertial reference frame $X_c$). 
The kinetic energy corresponding to the spin frequency $\omega$ is $E = h^* \omega = \gamma h^* \omega_0 = \gamma mc^2 \approx mc^2 + \frac{m}{2} \Vvec^2$ to first order in the small quantity 
$\frac{1}{c^2} \Vvec^2$ for a non-relativistic approximation. 

In the case of an electron in an external em-field, by substituting $E = \gamma mc^2$ and $\Pvec =  \gamma m \Vvec$ into the energy equation in Eq.~\eqref{Eq_1_9}, 
which includes the energy term $\Phi$ from the interaction between the spin and the external field, we get: 
\begin{equation}\label{Eq_1_10c}
	\gamma^2 = \left( 1 + \frac{1}{mc^2} \Phi \right) / \left( 1 - \frac{1}{c^2} \Vvec^2 \right)
\end{equation}
\noindent The spin frequency  $\omega = \gamma \omega_0$ is therefore affected 
by the spin-field interaction in such a way that the total kinetic energy $E = h^* \omega = \gamma h^* \omega_0 = \gamma mc^2 \approx mc^2 + \frac{m}{2} \Vvec^2 + \frac{1}{2} \Phi$ 
also includes a contribution from the interaction energy $\Phi$. 
The hypothesis that the spin-field interaction affects the time dilation $\gamma$ has been proposed by {\color{Green}van Holten (1992)}, although not explicitly through the flywheel analogy given here.

\subsection{Lagrangian Function for Spin Model}

Consider the Lagrangian function:
\begin{equation}\label{Eq_1_11}
	L = \frac{m} {2} u^{\mu} u_{\mu} + q A^{\mu}(x) u_{\mu} - \frac{m} {2 \omega_0^2} \dot{u}^{\mu} \dot{u}_{\mu}
\end{equation}
\noindent where $u = \dot{x}$ and $A = \left( \frac{V}{c}, \Avec \right)$ is the 4-vector potential for the em-field. 
This Lagrangian function is just the classical relativistic one with minimal coupling to the em-field (e.g. {\color{Green}Goldstein (1959)}) except for the additional last term, 
which is the contribution from the spin kinetic energy. For a free electron, $L$ is just the scaled sum of the two constraints $C1$ and $C2$ introduced in Eq.~\eqref{C1_C2}, 
where $C2$ implies that the last term in $L$ is just $\frac{m} {2} c^2$. 
For the classical relativistic Lagrangian function, $u = \dot{y}$ and it is the first term that is $\frac{m} {2} c^2$ whereas in $L$, the first term is zero by $C1$. 

The four Euler-Lagrange equations for stationarity of the action integral involving $L$ are:
\begin{equation}\label{Eq_1_12}
	\frac{d^2}{d \tau^2} \left(\frac{\del L}{\del \dot{u}_{\mu}} \right) - \frac{d}{d \tau} \left(\frac{\del L}{\del u_{\mu}} \right) + \frac{\del L}{\del x_{\mu}} = 0
\end{equation}
\noindent This produces a fourth-order equation for each of the four components of the total motion:
\begin{equation}\label{Eq_1_2}
	\ddddot{x\hspace{0pt}}^{\mu} + \omega_0^2 \ddot{x}^{\mu} - \frac{q \omega_0^2}{m} F^{\mu\nu} (x) \dot{x}_{\nu} = 0
\end{equation}
\noindent 	where, as usual, $F^{\mu \nu} = \frac{\del A^\nu }{\del x_{\mu}} - \frac{\del A^\mu }{\del x_{\nu}}$.
For complete specification of the solution, these equations of motion require the 16 initial conditions: $x(0)$, $\dot{x}(0)$, $\ddot{x}(0)$, $\dddot{x\hspace{0pt}}(0)$.

For each $\mu = 0,1,2,3$, the pair of second-order equations of motion in Eq.~\eqref{Eq_1_1} clearly imply the fourth-order equation in Eq.~\eqref{Eq_1_2} 
(just differentiate the first equation twice with respect to proper time). 
The converse is also seen to hold by setting $z^\mu = - \frac{1} {\omega_0^2}\ddot{x}^\mu$ and $y^\mu = x^\mu - z^\mu$, which gives immediately 
the first equation of Eq.~\eqref{Eq_1_1} and then the second equation follows by using $\ddot{y}^\mu = \ddot{x}^\mu - \ddot{z}^\mu$ in Eq.~\eqref{Eq_1_2}. 
Therefore, the fundamental equations of motion for the neo-classical model given in Eq.~\eqref{Eq_1_1} can be equivalently expressed as in Eq.~\eqref{Eq_1_2}. 

The Lagrangian function $L$ in Eq.~\eqref{Eq_1_11} is considered by {\color{Green}Riewe (1972)} for the free electron to derive a Hamiltonian function with spin that he used to perform a canonical quantization 
but the resulting equation was not directly related to Dirac's equation for the electron's wave function. 
{\color{Green}Salesi (2002)} also presents $L$ as  a special case of a more general Lagrangian function for a charged particle in an em-field defined by a scalar potential. 
He remarks that Eq.~\eqref{Eq_1_11} is of special interest for what he called a classical Dirac particle. His Hamiltonian function for this case is equivalent to that of {\color{Green}Riewe (1972)}.

\section{Spin Tensor Form of Equations of Spin Model}

\subsection{Spin Tensor Definition and Associated Equations}

We introduce a spin 4-tensor $S$ to describe the angular momentum of the \emph{total} momentum $mu = m \dot{x}$ of the electron about the spin center $C$ by defining its 16 components for $\mu , \nu = 0,1,2,3$ as:
\begin{equation}\label{Eq_1_13}
	S^{\mu\nu} = - m \left( z^{\mu}u^{\nu} - z^{\nu}u^{\mu} \right)
\end{equation}
\noindent By using Eq.~\eqref{Eq_1_1}, this new expression is equivalent to the definition given by {\color{Green}Salesi (2002)} who used N\"{o}ther's Theorem based on the invariance under infinitesimal 4-rotations 
of the Lagrangian function $L$ in Eq.~\eqref{Eq_1_11} with $A = 0$ to derive the spin tensor as: 
\begin{equation}\label{Eq_1_13b}
	S^{\mu\nu} = \frac{m} {\omega_0^2} \left( \dot{u}^{\mu}u^{\nu} - \dot{u}^{\nu}u^{\mu} \right)
\end{equation}

Based on Eq.~\eqref{Eq_1_13}, the proper time derivative of the spin tensor is $\dot{S}^{\mu \nu} = -m \left( \dot{z}^{\mu} u^{\nu} - u^{\mu} \dot{z}^{\nu} \right)$ because 
$z^{\mu} \dot{u}^{\nu} = \dot{u}^{\mu} z^{\nu}$ from Eq.~\eqref{Eq_1_1}. Thus, 
\begin{equation} \label{Eq_1_20}
	\dot{S}^{\mu \nu} = - \left[ \left( m u^{\mu} - \pi^{\mu} \right) u^{\nu} -  \left( m u^{\nu} - \pi^{\nu} \right) u^{\mu} \right] =  \pi^{\mu} u^{\nu} - \pi^{\nu} u^{\mu} 
\end{equation}
\noindent If we take the tensor for the orbital angular momentum about the origin O, $L^{\mu\nu} = x^{\mu}\pi^{\nu} - x^{\nu}\pi^{\mu}$, then:
\begin{equation}\label{Eq_1_20b}
	\dot{J}^{\mu\nu}  = \dot{L}^{\mu\nu} + \dot{S}^{\mu\nu} = x^{\mu}f^{\nu} - x^{\nu}f^{\mu} = M^{\mu\nu} 
\end{equation}
where $M^{\mu\nu}$ is the moment tensor about O of the em-force on the electron; in particular, the total angular momentum $J^{\mu\nu} = L^{\mu\nu} + S^{\mu\nu}$ is conserved for a free electron. 

Since $S$ is anti-symmetric ($S^{\nu\mu} = - S^{\mu\nu}$), the four diagonal components are zero and of the remaining 12 components, only 6 are independent, giving it the contravariant form:
\begin{equation}\label{Eq_1_14}
S = 
\left[  \begin{array}{cccc} 
0 & d^1 & d^2 & d^3 \\
-d^1 & 0 & -s^3 & s^2 \\
-d^2 & s^3 & 0 & -s^1 \\
-d^3 & -s^2 & s^1 & 0
\end{array} \right] 
\end{equation}
Here, $\svec$ $= \left(s^1, s^2, s^3 \right)$ is the spin vector and Eqs.~\eqref{Eq_1_13} and~\eqref{Eq_1_14} imply:
\begin{equation}\label{Eq_1_15}
	\svec = \zvec \times m \uvec 
\end{equation}
\noindent where $\uvec = \dot{\xvec}$, showing that $\svec$ is the 3-vector angular momentum for the total linear momentum $m\uvec$ about the spin center $C$. Differentiating and using Eq.~\eqref{Eq_1_1}, we get:
\begin{equation}\label{Eq_1_15b}
	\dot{\svec} = \uvec \times \Pvec
\end{equation}
\noindent which is just the spatial part of Eq.~\eqref{Eq_1_20}. It implies that the spin axis will precess if the observer is not fixed relative to the rest frame so that $\Pvec = 0$. 
Similarly, vector $\dvec = \left(d^1, d^2, d^3 \right)$ is given by:
\begin{equation}\label{Eq_1_16}
	\dvec = mc ( \dot{t} \zvec - t_z \uvec )
\end{equation}
\noindent Eq.~\eqref{Eq_1_20} implies that $\dot{\dvec} = \frac{1}{c} E \uvec - c \dot{t} \Pvec$.

Notice that the negative sign is needed out front in the definition of spin in Eq.~\eqref{Eq_1_13} if the total angular momentum is to be conserved for a free electron.
The chosen representation of the anti-symmetric tensor $S$ in terms of 3-vectors in Eq.~\eqref{Eq_1_14} then leads to the spin 3-vector $\svec$ as the angular momentum about the spin center $C$, 
as in Eq.~\eqref{Eq_1_15}. Using 3-vectors, the spatial part of Eq.~\eqref{Eq_1_20b} can now be written as:
\begin{equation}\label{Eq_1_14a}
	 \dot{\Jvec} = \frac{d}{d\tau} \left( \xvec \times \Pvec - \svec \right) =  \xvec \times {\bf f} 
\end{equation}
where $\Jvec = \xvec \times \Pvec - \svec = \yvec \times \Pvec - \zvec \times m\dot{\zvec}$ may be viewed as the 3-vector total angular momentum about O, which is conserved for a free electron. 
However, it is the difference between the 3-vector orbital angular momentum and the 3-vector spin, rather than the sum. This feature has been noted previously. 
For example, rather than using  Eq.~\eqref{Eq_1_15} to define the spin vector $\svec$, {\color{Green}Weyssenhoff (1947)} and {\color{Green}Rivas (2001)} define it to be the negative 
of the angular momentum of the total linear momentum. For the neo-classical spin model here, we follow instead the usual convention in classical mechanics that spin is an appropriate angular momentum 
(e.g. {\color{Green}Corben and Stehle (1960)}). 

It can be readily shown that the spin tensor $S$ satisfies two additional constraints: 
$\svec \cdot \svec = \left(\frac{\hbar}{2}\right)^2 \dot{t}^2$ and ${\bf d} \cdot {\bf d} = \left(\frac{\hbar}{2}\right)^2 \dot{t}^2$, so that $S$ has effectively only 4 independent components. 
(Notice that Eq.~\eqref{Eq_1_4b} implies that $\dot{t} = \dot{t}_y + \dot{t}_z$ has an ultra-high-frequency oscillation due to $\dot{t}_z$ about a temporal `average' (secular mean) 
$\dot{t}_y=\gamma$ when $\Pvec \ne 0$). These constraints give $S^{\mu \nu} S_{\mu \nu} = 2 \svec \cdot \svec - 2 {\bf d} \cdot {\bf d} = 0$. 
Furthermore, $\{ \frac{2}{\hbar \dot{t}} \dvec, \frac{1}{c \dot{t}}\uvec, \frac{2}{\hbar \dot{t}} \svec \}$ is a rotating right-handed orthonormal triad of 3-vectors because it is readily shown that:
\begin{equation}\label{Eq_1_16a}
	\text{(i) } \dvec = (\uvec \times \svec) / (c \dot{t}) \hspace{0.2in} \text{(ii) } \uvec = \frac{4c^2}{\hbar^2} (\svec \times \dvec) / (c \dot{t}) \hspace{0.2in} \text{(iii) } \svec =(\dvec \times \uvec ) / (c \dot{t})
\end{equation}

Using the definition of $S$ in Eq.~\eqref{Eq_1_13} and the first equation in Eq.~\eqref{Eq_1_1}, along with the constraints $C1$, $C2$ and $C3$ one at a time, 
the following identities for each $\mu = 0,1,2,3$ are readily proved using a line or two : 
\begin{align} \nonumber
	\text{(i) } & S^{\mu\nu} u_{\nu} = 0 \\ \nonumber
	\text{(ii) } & S^{\mu\nu} \dot{u}_{\nu} = mc^2 u^{\mu} \\ \nonumber
	\text{(iii) } & S^{\mu\nu} \pi_{\nu} = -(mc)^2 z^{\mu} \\  \nonumber
	\text{(iv) } & S^{\mu\nu} z_{\nu} = - \frac{\hbar}{2 \omega_0} u^{\mu} \\
	\text{(v) } & S^{\mu\nu} \dot{z}_{\nu} = mc^2 z^{\mu} \label{Eq_1_16b}
\end{align}
\noindent For example, (iii) follows from $S^{\mu \nu} \pi_{\nu} = -m \left( z^{\mu} u^{\nu}  \pi_{\nu} - u^{\mu} z^{\nu}  \pi_{\nu} \right) = - (mc)^2 z^{\mu}$ 
using $C3$ given in Eq.~\eqref{Eq_1_7} and its consequence in Eq.~\eqref{Eq_1_8}. 
The identity in (i) is often applied as a constraint when developing classical models of spin, although some authors impose $S^{\mu\nu} \pi_{\nu}  = 0$ instead, 
which does not hold in the neo-classical spin model presented here, as (iii) shows. 
It can be shown by using  Eq.~\eqref{Eq_1_1} and $C3$ that (i) holds if and only if $C1$ holds, that is, 
the electron moves at the speed of light. Eq.~\eqref{Eq_1_16a}(i) is just the spatial part of Eq.~\eqref{Eq_1_16b}(i). 


For a \emph{free} electron and using the rest-frame fixed at the spin center $C$ where $\Pvec = {\bf 0}$ and, as noted earlier, $\xvec = \zvec = \rvec, ~\uvec = \dot{\rvec}, ~t_z = 0$ and $t = \tau$, 
then (i) $\dvec = mc \rvec = \frac{\hbar}{2} {\bf e}_1$ where ${\rvec} = r_0 {\bf e}_1$, (ii) $\svec = \rvec \times m \dot{\rvec} = mcr_0 {\bf e}_1 \times {\bf e}_2 = \frac{\hbar}{2} {\bf e}_3$ where $\dot{\rvec} = c {\bf e}_2$, 
and (iii) $\dot{{\bf s}} = {\bf 0}$ so ${\bf e}_3$ is constant. Here, we have introduced $\{ {\bf e}_1, {\bf e}_2, {\bf e}_3 \}$ as a rotating right-handed orthonormal triad of 3-vectors, 
and used the consequences of the constraints $C1$ and $C2$. It is readily shown that $\dot{{\bf e}}_1 = \omega_0 {\bf e}_2 =$ \mbox{\boldmath $\Omega$} $\times {\bf e}_1$ where 
\mbox{\boldmath $\Omega$} $ = \omega_0  {\bf e}_3$ and $\dot{{\bf e}}_2 = - \omega_0 {\bf e}_1 =$ \mbox{\boldmath $\Omega$} $ \times {\bf e}_2$.
Thus, the circular spin motion takes place in the plane containing the orthogonal unit vectors ${\bf e}_1$ and ${\bf e}_2$, 
which is orthogonal to the constant spin direction ${\bf e}_3$ and it has a circular frequency of $\omega_0 = c/r_0$.

\subsection{Magnetic and Electric Dipole Energies}

The spin-field interaction energy $\Phi$ in Eq.~\eqref{Eq_1_9} can be expressed in terms of the spin and em-field tensors by using Eq.~\eqref{Eq_1_13}:
\begin{equation} \label{Eq_1_21}
	\Phi = z_{\mu} f^{\mu} = \frac{1}{2} \left(z_{\mu} q F^{\mu \nu} u_{\nu} - z_{\nu} q F^{\mu \nu} u_{\mu} \right) = - \frac{q}{2m} F^{\mu \nu}S_{\mu \nu}
\end{equation}
\noindent since $F^{\mu \nu} = - F^{\nu \mu}$. Substituting for the field and spin tensors: 
\begin{equation} \label{Eq_1_22}
	\Phi = - \frac{q}{m} \left( \Bvec \cdot \svec + \frac{1}{c} \Evec \cdot \dvec \right) = U_m + U_e
\end{equation}
\noindent showing that $\Phi$ is the sum of two energy terms, $U_m$ and $U_e$, due to magnetic and electric dipoles. Thus, the energy equation in Eq.~\eqref{Eq_1_9} becomes:
\begin{equation}\label{Eq_1_9b}
	\frac{1}{mc^2} E^2 = mc^2 + \frac{1}{m} \Pvec ^2 + U_m + U_e
\end{equation}
\noindent The dipole energies can be expressed as $U_m = - $ \mbox{\boldmath $\mu$} $\cdot \Bvec$ and $U_e = - $ \mbox{\boldmath $\epsilon$} $\cdot \Evec$ where 
\mbox{\boldmath $\mu$} $= \frac{q}{m} \svec$ and \mbox{\boldmath $\epsilon$} $= \frac{q}{mc} \dvec$ ($= q \rvec$ in the rest frame) are the orthogonal magnetic and 
electric moment 3-vectors, respectively. 
When the spin of an electron is aligned with a uniform magnetic field $\Bvec$, then $U_m =  \frac{\hbar}{2} \frac{e}{m} B = \frac{\hbar}{2} \omega_c$ 
where $\omega_c$ is the QM Larmor spin precession frequency.

Since $E=\gamma mc^2$ and $\Pvec =  \gamma m \Vvec$, the energy equation can be expressed as:
\begin{equation} \label{Eq_1_9c}
	E = \gamma^{-1} mc^2 + \Pvec \cdot \Vvec + \gamma^{-1} \Phi
\end{equation}
\noindent As shown in Eq.~\eqref{Eq_1_10c}, $\gamma$ depends on $\Phi$ and a first-order expansion in small quantities gives $\gamma^{-1} mc^2 \approx mc^2 - \frac{m}{2} \Vvec^2 - \frac{1}{2} \Phi$ 
and $\Pvec \cdot \Vvec \approx m \Vvec^2$, so Eq.~\eqref{Eq_1_9c} gives the approximation:  
\begin{equation} \label{Eq_1_9d}
	E \approx mc^2 + \frac{1}{2} m \Vvec^2 + \frac{1}{2} \Phi 
\end{equation}
\noindent agreeing with the result in Section 2.3.1. 
Thus, only a half of the spin-field interaction energy $\Phi$ contributes to the electron's total kinetic energy $E$, which an inertial observer can view as being stored 
in the spin motion, including spin precession (see the flywheel analogy in Section 2.3.1).
The apparent magnetic moment $\frac{1}{2}$\mbox{\boldmath$\mu$} agrees with classical electromagnetic theory if one argues that there is an approximate circular current loop created by the spin motion of the electron so that its magnetic moment has magnitude (loop current) $\times$ (loop area), giving 
$ \frac{q}{2\pi r_0/c} (\pi r_0^2) = \frac{q}{2m}(mc r_0) = \frac{q \hbar}{4m} = \frac{1}{2} \mu$. 


For a free electron, relative to the rest frame at C,  the spin vector $\svec =  \frac{\hbar}{2} {\bf e}_3$ is constant from Eq.~\eqref{Eq_1_15b} while the vector 
$\dvec = \frac{\hbar}{2} {\bf e}_1 = mc \rvec$ rotates at the ultra-high spin frequency $\omega_0$ due to the electron going around its circular path $\rvec = r_0 {\bf e}_1$ at speed $c$. 
In an electromagnetic field, an ultra-high spin frequency will still be present and so the electric dipole energy term in Eq.~\eqref{Eq_1_22} will have a temporal average (secular mean) of 
essentially zero in the rest frame for most electric fields. 
In an inertial observer frame where $\Pvec \ne {\bf 0}$, however, there will be a contribution from the electric dipole that does not average out to zero over 
small time intervals. To see this, apply Eq.~\eqref{Eq_1_16a}(i) to the electric dipole term in Eq.~\eqref{Eq_1_22}:
\begin{align} \nonumber
	U_e & = - \frac{q}{mc} \Evec \cdot \dvec = - \frac{q}{mc^2} \Evec \cdot (\uvec \times \svec)/\dot{t} = - \frac{q}{mc^2} (\Evec \times \uvec) \cdot \svec/\dot{t} \\ 
	& =  - \frac{q}{m^2 c^2} (\Evec \times \Pvec) \cdot \svec/\dot{t} - \frac{q}{mc^2} ( \Evec \times \dot{\zvec}) \cdot \svec/\dot{t}  \label{Eq_1_9e} 
\end{align} 
\noindent In general, the first term will dominate since the second term will have a secular mean of essentially zero for most electric fields because of the rotation of $\dot{\zvec}$ 
at ultra-high frequency. 
To within this approximaton, we note that the  non-relativistic approximation $\frac{1}{2} U_e$ for the electric dipole energy agrees with the corresponding operator term 
for the spin-orbit coupling in the corrected Pauli spin Hamiltonian in non-relativistic QM 
(e.g. Eq. (23-129) in {\color{Green} Baym (1981)} and {\color{Green} Deriglazov and Tereza (2019)}). 
This agreement of $\frac{1}{2} U_e$ is achieved without explicitly calling on Thomas precession ({\color{Green} Thomas (1926)}). However, the term in Pauli's Hamiltonian 
corresponding to the electron's magnetic dipole energy is $U_m$, which is twice that predicted for the non-relativistic case in the theory presented here. 

It is common practice in QM to follow {\color{Green} Thomas (1926)} and {\color{Green} Frenkel (1926)} and attribute the spin-orbit coupling term explaining the doublet structure 
of the spectral lines of the hydrogen atom to the change in the energy $U_m = - $ \mbox{\boldmath $\mu$} $\cdot \Bvec = - \frac{q}{m^2 c^2} \svec \cdot (\Evec \times \Pvec)$ 
under the induced magnetic field $\Bvec$ at the electron caused by its orbital motion $\Pvec$ in the proton's electric field $\Evec$ (e.g. {\color{Green} Spavieri and 
Mansuripur (2015)}). Mathematically, this gives the same expression as the dominant term for $U_e$ in Eq.~\eqref{Eq_1_9e} and so it is twice as large as it should be. 
For an explanation of the correct result, {\color{Green} Thomas (1926, 1927)} argued that the curved path of the electron induces a precession due to the successive boosts 
under the Lorentz transformation, giving a rotational kinetic energy that reduces the spin-orbit coupling energy by a half (e.g. Eq. (23-132) in {\color{Green} Baym (1981)}). 

{\color{Green} Dirac (1928, 1958)} in his analysis using the "square'' of his wave equation for the electron in an em-field, also gets Eqs.~\eqref{Eq_1_9} and~\eqref{Eq_1_21} 
in an operator form where his operator $\Phi$ corresponding to Eq.~\eqref{Eq_1_21} has the spin tensor $S$ replaced by its operator and it is a factor of two larger. 
Therefore, the magnetic and electric dipole energies are doubled, implying that in the non-relativistic case, these energies are $U_m$ and $U_e$, 
twice the predicted energies of the neo-classical theory here. 
Dirac's theory implies a magnetic moment that is close to experimental results for the electron. 
However, the electric dipole energy is twice as large as it should be if it is taken as the sole source of the separation of the spectral lines in the hydrogen atom spectrum. 
{\color{Green} Dirac (1928)} states: "The electric moment, being a pure imaginary, we should not expect to appear in the model. It is doubtful whether the electric moment 
has any physical meaning." 
The derivation of Eq.~\eqref{Eq_1_22} here gives a clear physical basis for the existence of both electric and magnetic dipole energies for the electron. 
This physical basis has also been noted by {\color{Green} Rivas (2001)} and {\color{Green} Hestenes (2010)}.

{\color{Green}Salesi and Recami (2000)} use the spin tensor form of the equations of motion from {\color{Green}Barut and Zanghi (1984)} to study the implied behavior 
of an electron in a constant uniform magnetic field. 
These equations of motion are given in Eq.~\eqref{Eq_1_18} in the next subsection where they are shown to be equivalent to those in Eq.~\eqref{Eq_1_1} under the 
constraint $C3$. However, the derivation of Salesi and Recami leads to an altered form of these equations because, in addition to $C3$, they use an analog of Dirac's operator 
form of the energy equation corresponding to Eq.~\eqref{Eq_1_9} where $\Phi$ in Eq.~\eqref{Eq_1_21} is doubled. 
This is not consistent with the equations of Barut and Zanghi because it implies that the quantity $G=z*\pi$ is nonzero since Eq.~\eqref{Eq_A_5} shows that $\dot{G} = - \Phi$, 
but it is proved in Appendix A that $G=0$ must hold. Because of this contradiction, the main conclusion of {\color{Green}Salesi and Recami (2000)} may not hold, namely, 
that the cyclotron frequency differs slightly for the two cases of the electron's spin being in the same or opposite direction to the magnetic field strength vector. 

{\color{Green}Hestenes (2010)} produces an energy equation like Eq.~\eqref{Eq_1_9b} for a classical model of the electron dynamics using Clifford space-time algebra by 
explicitly adding a dipole energy term to his Lagrangian function that has a Pauli spin-field interaction form. 
In contrast, in the neo-classical model presented here, the spin-field interaction terms exhibited in Eq.~\eqref{Eq_1_9b}  
arise solely from the minimal field-coupling term in the Lagrangian function in Eq.~\eqref{Eq_1_11}, which leads to the Lorentz force acting on the electron's spin motion. 
{\color{Green} Hestenes (2010)} also argues that there is experimental evidence ({\color{Green} Gouan\`ere et al. (2005)}) suggesting that the electron does indeed exhibit an 
ultra-fast rotating electric dipole due to its spin (``zitter'') motion and that it can cause resonance when an electron is propelled through a crystal ion channel at an appropriate speed. 

One possibility for doubling the spin-field interaction energy in Eq.~\eqref{Eq_1_21} so that it is consistent with Dirac's theory is to introduce further ``hidden" components of the spin motion. 
For example, {\color{Green}Consa (2018)} has suggested a toroidal solenoid spin model so for a free electron, the rest-frame circular spin motion presented here is replaced 
by a rest-frame toroidal motion of the electron moving at the speed of light. There are then two additional model parameters: the radius $r$ of the circular cross-section of the 
torus and the number $N$ of windings around the torus. 
For such a toroidal model, however, his speed of light constraint cannot be satisfied. The linear independence of the set of functions 
$\{1, \cos (N\omega \tau), \cos (2N\omega \tau) \}$ in the expression for the electron's speed (Eq. (36) of {\color{Green}Consa (2018)}) means that this constraint implies 
that $r=0$, so the toroidal spin model collapses to the circular one used in this work.

\subsection{Equations of Motion using Spin Tensor}

The eight second-order equations of motion in Eq.~\eqref{Eq_1_1} can be re-written as 16 equations in dynamic state-space form using the state $(x, u, y, \pi)$:
\begin{align} \nonumber
	\dot{x}^{\mu} & = u^{\mu} \\ \nonumber 
	\dot{u}^{\mu} & = - \omega_0^2 \left( x^{\mu} - y^{\mu} \right)  \\ \nonumber
	\dot{y}^{\mu} & = \frac{1}{m} \pi^{\mu}  \\ 
	\dot{\pi}^{\mu} & = q F^{\mu \nu} (x) u_{\nu} \label{Eq_1_17} 
\end{align}
\noindent Specification of the initial state at some time, make it $\tau = 0$, defines a unique solution to this set of equations for a specified field tensor $F$. 
They are linear except for the last equation, which is nonlinear unless there is no em-field or the electric and magnetic fields ${\bf E}(x)$ and ${\bf B}(x)$ are constant with respect to $x$. 

We now show that we can re-write these state-space equations in the following equivalent form using the state $(x, u, S, \pi)$:
\begin{align}  \nonumber
	\dot{x}^{\mu} & = u^{\mu}  \\ \nonumber
	\dot{u}^{\mu} & =  \frac{4c^2}{\hbar^2} S^{\mu \nu} \pi_{\nu}  \\ \nonumber
	\dot{S}^{\mu \nu} & = \pi^{\mu} u^{\nu} - \pi^{\nu} u^{\mu}  \\ 
	\dot{\pi}^{\mu} & = q F^{\mu \nu} (x) u_{\nu} \label{Eq_1_18}
\end{align}
\noindent The third equation is just Eq.~\eqref{Eq_1_20} that we proved using Eq.~\eqref{Eq_1_1}. The second equation comes from substituting (iii) of Eq.~\eqref{Eq_1_16b} into Eq.~\eqref{Eq_1_1}:
\begin{equation} \label{Eq_1_19}
	\dot{u}^{\mu} =  - \omega_0^2 z^{\mu} =   \frac{4c^2}{\hbar^2} S^{\mu \nu} \pi_{\nu} 
\end{equation}
\noindent This derivation shows that the equations of motion in Eq.~\eqref{Eq_1_1}, or equivalently Eq.~\eqref{Eq_1_17}, imply the equations of motion in Eq.~\eqref{Eq_1_18}. 
The converse is proved in Appendix A where the mapping between the initial conditions for~\eqref{Eq_1_17} and~\eqref{Eq_1_18} is also given.
Notice that the spin form of the set of equations of motion~\eqref{Eq_1_18} is nonlinear in the state except for a free electron where $\pi$ is a constant parameter. 
Also, the only property of the electron appearing explicitly in these equations is its charge $q=-e$.

Eq.~\eqref{Eq_1_18} reveals a connection of the neo-classical spin theory to Dirac's equation. These four equations are identical to the equations in the Appendix of 
{\color{Green}Grossmann and Peres (1963)} who derive them by replacing the commutator expression for the proper time derivatives of the operators representing dynamic 
quantities in the relativistic Heisenberg picture by an appropriate Poisson bracket for the dynamic quantities. 
They relegate these equations to an appendix because of reservations about their appropriateness. The equations that they present in the main body of their paper use 
observer time $t$ in the derivatives, rather than proper time $\tau$, and their sinusoidal dependence $\sin \omega t$ of the ZBW part of the free electron motion is not 
consistent with the $\sin \omega_0 \tau$ terms in Eq.~\eqref{Eq_1_4} except in the rest frame at $C$ where $\omega = \omega_0$ and $t = \tau$. 

{\color{Green}Krylovetskii (1978)} derives the Grossmann-Peres equations of motion using observer time $t$ by postulating a Lagrangian function involving a neo-classical 
spinor of the form investigated by {\color{Green}Proca (1954)}. 
He then shows that these equations using observer time $t$ are indeed covariant when moving from one inertial reference frame to another. 
{\color{Green}Barut and Zanghi (1984)} use a proper time version of Krylovetskii's spinor Lagrangian function, although apparently unaware of his work, and show that the 
resulting equations imply Eq.~\eqref{Eq_1_18}. We return to their theory in the next section. 

{\color{Green}Hestenes (2010)} presents a similar Lagrangian function involving spinors in proper time but formulated using Clifford space-time algebra. 
His derived equations of motion are equivalent to those in Eq.~\eqref{Eq_1_18} if the explicit dipole energy term that he adds to his Lagrangian function is dropped. 
As we showed in the previous subsection, the Lagrangian function in Eq.~\eqref{Eq_1_11} has implicitly electric and magnetic dipole moments because of the action of 
the Lorentz force on the spin motion arising from its spin energy term. 

{\color{Green}Barut (1987)} shows that the relativistic Heisenberg operator equations for the proper-time Dirac equation have a correspondence with the spin tensor form in 
Eq.~\eqref{Eq_1_18} where the dynamic variables in the state correspond to their operator representations. 
A similar correspondence is exhibited later in the operator equations in Eq.~\eqref{Eq_1_34}. 

Finally, {\color{Green}Weyssenhoff and Raabe (1947)} develop equations of motion for an electron moving at the speed of light that are very similar to Eq.~\eqref{Eq_1_18} 
except that they have an implicit equation for $\dot{u}$ rather than the explicit one in Eq.~\eqref{Eq_1_18}. 
They derive their equations by applying linear and angular momentum conservation to a fluid that has volume elements that spin, 
then they let the elements shrink to a point in the limit of an infinitely large mass density and angular momentum density. Their paper, and the others just referenced, 
show that there is something special about the equations in Eq.~\eqref{Eq_1_18}, which are apparently applicable to any fundamental charged particle with spin $ \hbar/2$.

\section{Operator Form of Equations of Spin Model}

\subsection{Neo-classical Dirac-Schr\"{o}dinger Equation}

It is shown in this section that the spin tensor form of the equations of motion can be represented using operators on a four-dimensional complex state function 
that is a solution in proper time of a neo-classical Dirac-Schr\"{o}dinger equation. 

Based on {\color{Green}Barut and Zanghi (1984)}, we introduce a \emph{state function} $\phi(\tau) \in {\cal C}^4$, a 4-dimensional complex column vector function, satisfying the neo-classical spinor Dirac-Schr\"{o}dinger equation:
\begin{equation} \label{Eq_1_23}
	 i \hbar \dot{\phi} = \hat{H} \phi
\end{equation}
\noindent where the $4 \times 4$ complex matrix $\hat{H} = c \pi_{\mu} \gamma^{\mu}$ is the Hamiltonian operator and $\gamma^{\mu}, \mu = 0,1,2,3,$ are the usual $4 \times 4$ matrices 
in the covariant form of Dirac's wave equation ({\color{Green}Bjorken and Drell (1964)}). 
Notice that this spinor equation represents 8 real-valued equations involving the real and imaginary parts of the complex column vector $\phi$. 
As shown later, for a free electron the state function $\phi$ corresponds to Dirac's wave function after a Lorentz transformation. 
Eq.~\eqref{Eq_1_23} shows that $\phi$ depends on the electron's motion only through its global momentum $\pi$.

As in Dirac's theory, we define the $4 \times 4$ constant matrix operators:
\begin{equation} \label{Eq_1_24}
\hat{u}^{\mu} = c \gamma^{\mu}, \hspace{0.2in} \hat{S}^{\mu\nu} = - \frac{i \hbar}{4} \left[ \gamma^{\mu}\gamma^{\nu} - \gamma^{\nu}\gamma^{\mu} \right] 
\end{equation}
\noindent that act on $\phi$. This notation for these two operators anticipates the results of Section 4.2 where it is shown that the total velocity component $u^{\mu} = \bar{\phi} \hat{u}^{\mu} \phi$ and 
the spin tensor component $S^{\mu\nu} = \bar{\phi} \hat{S}^{\mu\nu} \phi$ give the equations of motion in Eq.~\eqref{Eq_1_18}, where, as in QM theory, we define $\bar{\phi} = \phi^{*} \gamma^0$ 
with $\phi^{*}$ the Hermitian (conjugate) transpose of $\phi$. 
The negative sign at the front of the expression for $\hat{S}$ is needed for consistency with $S$ in Eq.~\eqref{Eq_1_13}, which gives conservation of total angular momentum about the origin $O$ for the free electron. 
Notice also that the matrix operators $\hat{u}^{\mu}$ and $\hat{S}^{\mu\nu}$ are not Hermitian but $\gamma_0 \hat{u}^{\mu}$ and $\gamma_0 \hat{S}^{\mu\nu}$ are, which are the actual operators 
in the inner product $\phi^{*} (.) \phi$ that defines the usual Hilbert space on ${\cal C}^4$. The time operator $\hat{u}^0 = c \gamma^0$ gives $c \dot{t} = u^0 = \bar{\phi} \hat{u}^0 \phi = c \phi^* \phi$.
Thus, as noted after Eq.~\eqref{Eq_1_4b}, observer time $t$ increases monotonically with proper time $\tau$ because $\dot{t} = \phi^* \phi$ is always positive.

If $\hat{Q}(\tau)$ is an operator and $Q(\tau)$ is the corresponding dynamical variable defined by the inner product $Q = \phi^{*} (\gamma^0 \hat{Q}) \phi = \bar{\phi} \hat{Q} \phi$, then the derivative of $Q(\tau)$ is given by: 
\begin{align}  \nonumber
	 \dot{Q} & = \dot{\bar{\phi}} \hat{Q} \phi + \bar{\phi} \hat{Q} \dot{\phi} + \bar{\phi} \dot{\hat{Q}} \phi \\
	             & = \frac{i}{\hbar}  \bar{\phi} \left( \hat{H}\hat{Q}  - \hat{Q} \hat{H}  \right)  \phi + \bar{\phi} \dot{\hat{Q}} \phi \label{Eq_1_27}
\end{align}
\noindent where we use Eq.~\eqref{Eq_1_23} in the first two terms. This proves a result that is familiar from QM for the time derivative of an observable in terms of a commutator with the Hamiltonian operator:
\begin{equation} \label{Eq_1_28}
	\hat{\dot{Q}} = \frac{i}{\hbar} \left[ \hat{H}, \hat{Q} \right] + \dot{\hat{Q}}
\end{equation}
\noindent where from Eq.~\eqref{Eq_1_27}, $\dot{Q} = \bar{\phi} \hat{\dot{Q}} \phi$. Note that $\hat{H}$ could be time-varying because $\pi$ can be if the electron is in an em-field 
but the Hamiltonian function $H$ corresponding to $\hat{H}$ is constant since:
\begin{equation} \label{Eq_1_29}
	\dot{H} = \bar{\phi} \hat{\dot{H}} \phi =  \bar{\phi} \frac{i}{\hbar} \left[ \hat{H}, \hat{H} \right] \phi +  \bar{\phi} \dot{\hat{H}} \phi = \dot{\pi}_{\mu}u^{\mu} = 0
\end{equation}
\noindent by using the definition of $\hat{H}$ and $\hat{u}^{\mu}$ above and then using Eq.~\eqref{Eq_1_17} for $\dot{\pi}$ in terms of the anti-symmetric tensor $F$. 
The time invariance of $H$ can also be shown directly:
\begin{equation} \label{Eq_1_30}
	H = \bar{\phi} \hat{H} \phi = \pi_{\mu} \bar{\phi} \hat{u}^{\mu} \phi = \pi_{\mu} u^{\mu} = mc^2
\end{equation}
\noindent using constraint $C3$. We use this equation to provide a normalization of the solution $\phi$ of the homogeneous linear equation~\eqref{Eq_1_23}.

\subsection{Operator Form of Equations of Motion}

Consider the following state-space equations using the state $(x, \phi, \pi)$:
\begin{align}  \nonumber
	\dot{x}^{\mu} & = \bar{\phi} \hat{u}^{\mu} \phi  \\ \nonumber
	\dot{\phi} & = - \frac{i}{\hbar} \hat{H} \phi = - \frac{ic}{\hbar} \pi_{\nu}  \gamma^{\nu} \phi   \\
	\dot{\pi}^{\mu} & = q F^{\mu\nu}(x) \bar{\phi} \hat{u}_{\nu} \phi   \label{Eq_1_31}
\end{align}
\noindent We now show that a solution of these equations (equivalent to 16 real-valued equations) gives the solution of the equations of motion in Eq.~\eqref{Eq_1_18} with 
initial conditions: $x(0)$, $\pi(0)$, $u(0) = \bar{\phi}(0) \hat{u} \phi(0)$ and $S(0) = \bar{\phi}(0) \hat{S} \phi(0)$. 
The state function $ \phi (\tau)$, in conjunction with Newton's Second Law, therefore gives an alternative but equivalent way to describe the electron's dynamics in the neo-classical model. 

First, since $u^{\mu} = \dot{x}^{\mu} = \bar{\phi} \hat{u}^{\mu} \phi$, the first and last equations in Eq.~\eqref{Eq_1_31} are identical to those of Eq.~\eqref{Eq_1_18}. 
Furthermore, $\dot{u}^{\mu} = \bar{\phi} \hat{\dot{u}}^{\mu} \phi$ where from Eq.~\eqref{Eq_1_28}:
\begin{align}  \nonumber
	\hat{\dot{u}}^{\mu} & = \frac{i}{\hbar} \left[ \hat{H}, \hat{u}^{\mu} \right]  \\ \nonumber
	                      & = \frac{ic}{\hbar} \left[ \hat{H}, \gamma^{\mu} \right]  \\ \nonumber
	                      & = \frac{ic^2}{\hbar} \left[\pi_{\nu}  \gamma^{\nu}  \gamma^{\mu} - \pi_{\nu}  \gamma^{\mu}  \gamma^{\nu} \right]  \\
	                      & = \frac{4c^2}{\hbar^2} \hat{S}^{\mu\nu} \pi_{\nu}  \label{Eq_1_32}
\end{align}
\noindent Thus, $\dot{u}^{\mu} = \bar{\phi} \hat{\dot{u}}^{\mu} \phi = \frac{4c^2}{\hbar^2} \bar{\phi} \hat{S}^{\mu\nu} \phi \pi_{\nu} = \frac{4c^2}{\hbar^2} S^{\mu\nu} \pi_{\nu}$, as in Eq.~\eqref{Eq_1_18}. 

Second, consider $\dot{S}^{\mu\nu} = \bar{\phi} \hat{\dot{S}}^{\mu\nu} \phi$. The Dirac matrices satisfy the identity $\gamma^{\mu}\gamma^{\nu} + \gamma^{\nu}\gamma^{\mu} = 2 g^{\mu\nu} I_4$ 
where $I_4$ is the $4 \times 4$ identity matrix and so $\hat{S}^{\mu\nu} = - \frac{i \hbar}{2} \left[ \gamma^{\mu}\gamma^{\nu} - g^{\mu\nu} I_4 \right]$. 
If $\mu = \nu$, then $\gamma^\mu \gamma^\mu = g^{\mu\mu} I_4$ and $\hat{S}^{\mu\nu} = 0$. 
If $\mu \neq \nu$, then $\gamma^\mu \gamma^\nu = - \gamma^\nu \gamma^\mu$ and $\hat{S}^{\mu\nu} = - \frac{i\hbar}{2} \gamma^\mu \gamma^\nu$, so from Eq.~\eqref{Eq_1_28}: 
\begin{align}  \nonumber
	\hat{\dot{S}}^{\mu\nu} & = \frac{i}{\hbar} \left[ \hat{H}, \hat{S}^{\mu\nu} \right]  \\ \nonumber
	                           & = \frac{1}{2} \left[ \hat{H}, \gamma^{\mu} \gamma^{\nu} \right]  \\ \nonumber
	                           & = \frac{c}{2} \left[ \pi_{\sigma} \gamma^{\sigma}\gamma^{\mu}\gamma^{\nu} - \pi_{\sigma} \gamma^{\mu}\gamma^{\nu}\gamma^{\sigma} \right]  \nonumber
\end{align}
\noindent But if $\sigma \neq \mu, \nu$, then $\gamma^\sigma \gamma^\mu \gamma^\nu = - \gamma^\mu \gamma^\sigma \gamma^\nu = \gamma^\mu \gamma^\nu \gamma^\sigma$, 
so only the terms with $\sigma = \mu,\nu$ appear in the sum over $\sigma$, implying that: 
\begin{align}  \nonumber
	\hat{\dot{S}}^{\mu\nu} & = \frac{c}{2} \pi_\mu \left[ \gamma^\mu \gamma^\mu \gamma^\nu - \gamma^\mu \gamma^\nu \gamma^\mu \right]  
						+ \frac{c}{2} \pi_\nu \left[ \gamma^\nu \gamma^\mu \gamma^\nu - \gamma^\mu \gamma^\nu \gamma^\nu \right]  \\ \nonumber
	                           & = c \pi_{\mu} g^{\mu\mu} \gamma^{\nu} - c \pi_{\nu} g^{\nu\nu} \gamma^{\mu}  \\ 
	                           & = \pi^{\mu}\hat{u}^{\nu} - \pi^{\nu}\hat{u}^{\mu}   \label{Eq_1_33}
\end{align}
\noindent Thus, $\dot{S}^{\mu\nu} = \bar{\phi} \hat{\dot{S}}^{\mu\nu} \phi = \pi^{\mu} u^{\nu} - \pi^{\nu} u^{\mu}$, as in Eq.~\eqref{Eq_1_18}. 
In effect, the neo-classical Dirac-Schr\"{o}dinger equation for the evolution of the spinor state function $\phi$ in Eq.~\eqref{Eq_1_31} replaces the evolution equations for the total 4-velocity $u$ 
and the spin tensor $S$ in Eq.~\eqref{Eq_1_18}, so that the two sets of state-space equations produce the same solution for all dynamic variables. 

We note that Eq.~\eqref{Eq_1_18} can now be equivalently expressed as operator equations:
\begin{align}  \nonumber
	\hat{\dot{x}}^{\mu} & = \hat{u}^{\mu}  \\ \nonumber
	\hat{\dot{u}}^{\mu} & =  \frac{4c^2}{\hbar^2} \hat{S}^{\mu \nu} \pi_{\nu}  \\ \nonumber
	\hat{\dot{S}}^{\mu \nu} & = \pi^{\mu} \hat{u}^{\nu} - \pi^{\nu} \hat{u}^{\mu}  \\ 
	\hat{\dot{\pi}}_{\mu} & = q F_{\mu \nu} (x) \hat{u}^{\nu} \label{Eq_1_34}
\end{align}
\noindent where a dynamic variable $Q(\tau) = \bar{\phi} \hat{Q} \phi$ and $\phi(\tau)$ is given by Eq.~\eqref{Eq_1_23} with $\phi(0)$ specified. 
These equations are identical to the Heisenberg equations from Dirac's equation in proper time ({\color{Green}Barut (1987)}), except that the actual global momentum $\pi$ is involved, 
not an operator for it. We introduce an operator for $\pi$ for the neo-classical theory in a later subsection where a wave function is defined in terms of the state function.

\subsection{Superposition of State Functions for a Free Electron}

For the free electron, $\hat{H}$ in Eq.~\eqref{Eq_1_23} is not a function of $\tau$, so:
\begin{equation} \label{Eq_1_41}
	 \ddot{\phi} = \frac{1}{i \hbar} \frac{d}{d\tau} \left( \hat{H} \phi \right) =   - \frac{1}{\hbar^2} \hat{H} \left(i \hbar \dot{\phi} \right) = - \frac{1}{\hbar^2} \hat{H}^2 \phi = - \omega_1 ^2 \phi
\end{equation}
\noindent setting $\omega_1 = \frac{mc^2}{\hbar} = \frac{1}{2} \omega_0$ after using the identity:
\begin{equation} \label{Eq_1_39}
	\hat{H}^2 = c^2 \pi_{\mu} \pi_{\nu} \gamma^{\mu} \gamma^{\nu} = \frac{1} {2} c^2 \pi_{\mu} \pi_{\nu} \left( \gamma^{\nu} \gamma^{\mu} + \gamma^{\mu} \gamma^{\nu} \right) = c^2 \pi_{\mu} \pi_{\nu} g^{\nu\mu} I_4 = (mc^2)^2  I_4
\end{equation}
\noindent where the last equality is from Eq.~\eqref{Eq_1_9} with $\Phi = 0$. This simple linear oscillator equation is a neo-classical spin theory version of the \emph{Klein-Gordon equation} in QM and 
it has the solution:
\begin{align}  \nonumber
	\phi(\tau) & = \cos (\omega_1 \tau) \phi(0) + \frac{1}{\omega_1} \sin (\omega_1 \tau) \dot{\phi}(0) \\
	               & = \left[ \cos (\omega_1 \tau) I_4 - \frac{i}{mc^2} \sin (\omega_1 \tau) \hat{H} \right] \phi(0) \label{Eq_1_42}
\end{align}
\noindent using Eq.~\eqref{Eq_1_23} in the second equation.
This solution of Eq.~\eqref{Eq_1_41} is readily shown to also be the solution of Eq.~\eqref{Eq_1_23}. 
It was first given by {\color{Green}Barut and Zanghi (1984)} in their Eq.\  (4) along with an expression for $u^{\mu} = \bar{\phi} \hat{u}^{\mu} \phi$ (see Appendix B for a detailed derivation), 
which agrees with $\dot{x} (\tau)$ in Eq.~\eqref{Eq_1_4}, as expected. (Their Eq.\  (4) has three typographical errors). 
It is readily shown that if $\phi(0)$ satisfies the energy normalization in Eq.~\eqref{Eq_1_30}, so does $\phi(\tau)$ in Eq.~\eqref{Eq_1_42} for all proper time $\tau$.

Expressing the sine and cosine terms in Eq.~\eqref{Eq_1_42} as complex exponentials:
\begin{equation} \label{Eq_1_45}
	\phi(\tau) = \left[ \frac{1}{2} \left(I_4 + \frac{1}{mc^2} \hat{H} \right) \exp (-i \omega_1 \tau) + \frac{1}{2} \left(I_4 - \frac{1}{mc^2} \hat{H} \right) \exp (i \omega_1 \tau) \right] \phi(0)
\end{equation}
\noindent For arbitrary $\phi(0) = A \in {\cal C}^4$ satisfying the energy normalization $\bar{A} \hat{H} A = mc^2$ from Eq.~\eqref{Eq_1_30}, define:
\begin{equation} \label{Eq_1_45a}
	A^+ = \frac{1}{2} \left(I_4 + \frac{1}{mc^2} \hat{H} \right) A, \hspace{0.2in} A^- = \frac{1}{2} \left(I_4 - \frac{1}{mc^2} \hat{H} \right) A
\end{equation}
\noindent then using the identity~\eqref{Eq_1_39}, $\hat{H} A^+ = mc^2 A^+$ and $\hat{H} A^- = - mc^2 A^-$. Thus, $A^+$ and $A^-$ are eigenvectors of $\hat{H}$ belonging 
to the two-dimensional eigenspaces corresponding to eigenvalues $mc^2$ and $-mc^2$, 
respectively, that depend on the global momentum $\pi$. From Eq.~\eqref{Eq_1_45}, any solution of Eq.~\eqref{Eq_1_23} for the free electron is a superposition:
\begin{equation} \label{Eq_1_46a}
	\phi (\tau) = \phi^+ (\tau) + \phi^- (\tau), \hspace{0.2in} \phi^+ (\tau) = \exp (- i\omega_1 \tau) A^+,  \hspace{0.2in} \phi^- (\tau) = \exp ( i\omega_1 \tau) A^-
\end{equation}
\noindent where these special state function solutions satisfy $\hat{H} \phi^+ = mc^2 \phi^+$ and $\hat{H} \phi^- = - mc^2 \phi^-$. 
They correspond to the so-called \emph{positive energy} and \emph{negative energy} solutions, respectively, in relativistic QM theory but this is misleading terminology here. 
As stand-alone solutions of Eq.~\eqref{Eq_1_23}, the normalization in~\eqref{Eq_1_30} implies that $\phi^+$ and $\phi^-$ both have positive energy, as expected 
for a neo-classical relativistic mechanics theory, and $\bar{A}^+ A^+ = 1$ and $\bar{A}^- A^- = -1$. In general, Eq.~\eqref{Eq_1_45a} imply that 
$\bar{A}^+ A^+ = \frac{1}{2}(\bar{A} A + 1), \bar{A}^- A^- = \frac{1}{2}(\bar{A} A - 1)$ and $\bar{A}^+ A^-, \bar{A}^- A^+$ are both zero. 

To investigate spin states, consider the special case where the observer's inertial reference frame coincides with the rest frame fixed at the spin center C, then $\hat{H} = mc^2 \gamma^0$ and 
so the eigenvectors of $\hat{H}$ have the form:
\begin{equation} \label{Eq_1_47a}
	A^+ = \left[ A_1, A_2, 0,0 \right]^T, \hspace{0.2in} A^- = \left[ 0,0, A_3, A_4 \right]^T
\end{equation}
\noindent From Eq.~\eqref{Eq_1_46a}, we can express the general solution of Eq.~\eqref{Eq_1_23} for the free electron case as the superposition:
\begin{equation} \label{Eq_1_48a}
	\phi(\tau) = \phi^+ (\tau) + \phi^- (\tau) = \left[ A_1 e^{-i \omega_1 \tau}, A_2 e^{-i \omega_1 \tau}, A_3 e^{i \omega_1 \tau}, A_4 e^{i \omega_1 \tau} \right]^T
\end{equation}
\noindent The complex amplitudes $A_i = \phi_i(0)$ must satisfy the previous normalization condition that gives positive energy $mc^2$, so $mc^2 = \bar{\phi} \hat{H} \phi = mc^2 \phi^* \phi$, which implies: 
\begin{equation} \label{Eq_1_49a}
	1 = \phi^* \phi = |A_1|^2 + |A_2|^2 + |A_3|^2 + |A_4|^2
\end{equation}
\noindent This is consistent with the requirement from Eq.~\eqref{Eq_1_24} that any state function must satisfy $\phi^*  \phi = \frac{1}{c} \bar{\phi} \hat{u}^0 \phi = \frac{1}{c} u^0 = \dot{t} =  1$ 
because $t=\tau$ in the rest frame at $C$. 

In Appendix C, the total velocity components $u^j$ and spin components $s^j$ $(j = 1,2,3)$ are given for the state function $\phi(\tau)$ in Eq.~\eqref{Eq_1_48a}. 
Consider the choice $A_1 = \frac{1}{\sqrt{2}} \exp (-i \varphi/2), A_2 = 0, A_3 = 0, A_4 = \frac{1}{\sqrt{2}} \exp (i \varphi/2)$, 
then the normalization in Eq.~\eqref{Eq_1_49a} is satisfied and from Appendix C:
\begin{align}  \nonumber
	u^1 & = c \cos (\omega_0 \tau + \varphi), \hspace{0.1in} u^2 = c \sin (\omega_0 \tau + \varphi), \hspace{0.1in} u^3 = 0 \\
	s^1 & = s^2 = 0, \hspace{0.3in} s^3 = \hbar/2  \label{Eq_1_50a}
\end{align}
\noindent This choice for the state function $\phi$ corresponds to spin-up in the direction of the $x^3$ axis where the electron is moving in a circle in the $x^1-x^2$ plane at speed $c$ 
with angular momentum $\hbar/2$ and so radius $r_0 = \frac{\hbar}{2mc}$. 
Now consider the choice $A_1 = 0, A_2 = = \frac{1}{\sqrt{2}} \exp (i \varphi/2), A_3 = - \frac{1}{\sqrt{2}} \exp (-i \varphi/2), A_4 = 0$ in  Eq.~\eqref{Eq_1_48a}, then:
\begin{align}  \nonumber
	u^1 & = - c \cos (\omega_0 \tau - \varphi), \hspace{0.1in} u^2 = c \sin (\omega_0 \tau - \varphi), \hspace{0.1in} u^3 = 0 \\
	s^1 & = s^2 = 0, \hspace{0.3in} s^3 = - \hbar/2  \label{Eq_1_51a}
\end{align}
\noindent This choice for $\phi$ corresponds to spin-down for the $x^3$ direction. 
The solutions in Eqs.~\eqref{Eq_1_50a} and~\eqref{Eq_1_51a} agree with those given in Subsection 2.2 for the initial phase $\varphi_0 = \varphi - \pi/2$. 
Notice also that the spin-up and spin-down state functions here are both linear superpositions of the ``positive energy" and ``negative energy" solutions in Eq.~\eqref{Eq_1_46a}.

Consider the electron spinning about an axis defined by a unit vector $\nvec = (n^1,n^2,n^3) = n^j \evec_j$ (summation convention over $j=1,2,3$) where $\{\evec_1, \evec_2, \evec_3 \}$ is an orthonormal triad of 3-vectors. 
The spin vector is then $\svec_n = s_n \nvec$ where $s_n = \svec_n \cdot \nvec = (s^j \evec_j) \cdot \nvec = s^j n^j$.
Using the operator for each $s^j$ given in Appendix C, the operator for $s_n$ is therefore: 
\begin{equation} \label{Eq_1_52a}
	\hat{s}_n = n^j \hat{s}^j = \frac{\hbar}{2} \left[  \begin{array}{cc} 
		\sigma_n & 0 \\
		0 & \sigma_n 
\end{array} \right] 
\end{equation}
\noindent where $\sigma_n = n^j \sigma^j$ and the $\sigma^j$'s are the three Pauli spin matrices. Using spherical coordinates with polar angle $\theta$ and the $x^3$-axis as the polar axis:
\begin{equation} \label{Eq_1_53a}
	n^1 = \sin \theta \cos \varphi, \hspace{0.1in} n^2 = \sin \theta \sin \varphi, \hspace{0.1in} n^3 = \cos \theta 
\end{equation}
\noindent and so 
\begin{equation} \label{Eq_1_54a}
	\sigma_n = \left[  \begin{array}{cc} 
		\cos \theta & \exp (-i \varphi) \sin \theta \\
		\exp (i \varphi) \sin \theta & - \cos \theta 
\end{array} \right] 
\end{equation}

For any state $\phi (\tau)$, $s_n = \bar{\phi} \hat{s}_n \phi$, so:
\begin{equation} \label{Eq_1_55a}
	s_n = \frac{\hbar}{2} \phi^* \Sigma_n \phi = 
		\frac{\hbar}{2} \phi^* 
		\left[  \begin{array}{cc} 
		\sigma_n & 0 \\
		0 & - \sigma_n  
	\end{array} \right] \phi
\end{equation}
\noindent where $\phi^* \phi = 1$ and $\Sigma_n$ is a Hermitian matrix. 
By Rayleigh's Quotient Theorem, $s_n$ has its maximum value $\frac{\hbar}{2}$ at the maximum eigenvalue $\lambda = 1$ of $\Sigma_n$ and its minimum value 
$- \frac{\hbar}{2}$ at the minimum eigenvalue $\lambda = -1$. These values occur at the corresponding normalized eigenvectors of $\Sigma_n$. 
This implies that the state function $\phi_n (\tau) \in {\cal C}^4$ that gives the spin vector $\svec_n = s_n \nvec =  \frac{\hbar}{2} \nvec$ 
is the eigenvector corresponding to eigenvalue $\lambda = 1$, which is given by the special case of Eq.~\eqref{Eq_1_48a} with coefficients:
\begin{align}  \nonumber
	A_1 & = \frac{1}{\sqrt{2}} \exp (-i \varphi/2) \cos (\theta/2), \hspace{0.1in} & A_2 = \frac{1}{\sqrt{2}} \exp (i \varphi/2) \sin (\theta/2) \\
	A_3 & = - \frac{1}{\sqrt{2}} \exp (-i \varphi/2) \sin (\theta/2), \hspace{0.1in} & A_4 = \frac{1}{\sqrt{2}} \exp (i \varphi/2) \cos (\theta/2)  \label{Eq_1_56a}
\end{align}
\noindent (Because of the structure of $\Sigma_n$, this eigenvector has as its first two and last two components equal to the two spin-up and two spin-down components, respectively, from the non-relativistic Pauli spin theory). 

Substituting these coefficients into the expressions in Appendix C, the velocity components and spin components corresponding to the spin state $\phi_n (\tau)$ are:
\begin{align}  \nonumber
	u^1 & = c \cos^2 \frac{\theta}{2} \cos(\omega_0 \tau + \varphi) - c \sin^2 \frac{\theta}{2} \cos(\omega_0 \tau - \varphi)  \\  \nonumber
		& = c \cos \theta \cos \varphi \cos (\omega_0 \tau) - c \sin \varphi \sin (\omega_0 \tau) \\  \nonumber
	u^2 & = c \cos^2 \frac{\theta}{2} \sin(\omega_0 \tau + \varphi) + c \sin^2 \frac{\theta}{2} \sin(\omega_0 \tau - \varphi)  \\  \nonumber
		& = c \cos \varphi \sin (\omega_0 \tau) + c \cos \theta \sin \varphi \cos (\omega_0 \tau) \\
	u^3 & =  - c \sin \theta  \cos (\omega_0 \tau)  \label{Eq_1_57a}
\end{align}
\begin{align}  \nonumber
	s^1 & =  \frac{\hbar}{2} \sin \theta \cos \varphi = \frac{\hbar}{2} n^1  \\  \nonumber
	s^2 & =  \frac{\hbar}{2} \sin \theta \sin \varphi  = \frac{\hbar}{2} n^2  \\
	s^3 & =  \frac{\hbar}{2} \cos \theta = \frac{\hbar}{2} n^3  \label{Eq_1_57b}
\end{align}
As expected, the $s^j$ are the projections of the spin vector $\svec_n =  \frac{\hbar}{2} \nvec$ onto the axes defined by the orthonormal triad of 3-vectors $\{\evec_1, \evec_2, \evec_3 \}$. 

For the special cases $\theta = 0$ and $\theta = \pi$, Eqs.~\eqref{Eq_1_57a} and ~\eqref{Eq_1_57b} give agreement with the velocity and spin components for the spin-up and spin-down states 
in Eqs.~\eqref{Eq_1_50a} and~\eqref{Eq_1_51a}, respectively, as expected. 
The corresponding state functions $\phi_{up}$ and $\phi_{dn}$ are given by substituting $\theta = 0$ and $\theta = \pi$ in the coefficients in Eq.~\eqref{Eq_1_56a} and using them in Eq.~\eqref{Eq_1_48a}:
\begin{align}  \nonumber
	\phi_{up} & = \left[ \frac{1}{\sqrt{2}} \exp (-i (\omega_0 \tau + \varphi)/2), 0, 0, \frac{1}{\sqrt{2}} \exp (i (\omega_0 \tau + \varphi)/2) \right]^T \\
	\phi_{dn} & = \left[ 0, \frac{1}{\sqrt{2}} \exp (-i (\omega_0 \tau - \varphi)/2), - \frac{1}{\sqrt{2}} \exp (i (\omega_0 \tau - \varphi)/2), 0 \right]^T \label{Eq_1_58a}
\end{align}
The state function $\phi_n (\tau)$ for spin about direction $\nvec$ can therefore be written as a superposition of $\phi_{up}$ and $\phi_{dn}$, which are orthogonal unit vectors under the usual Hermitian inner product on ${\cal C}^4$. 
Furthermore, using Euler's formula for $\exp (i \varphi/2)$, it is readily seen that $\phi_{up}$ and $\phi_{dn}$ are themselves linear superpositions of ``positive energy" and ``negative energy" solutions, which are 
the orthogonal unit vectors given by setting $\varphi = 0$ and $\varphi = \pi$ in each of Eq.~\eqref{Eq_1_58a}. Therefore, the spin state for direction $\nvec$ is a linear superposition of spin-up and spin-down states as follows:
\begin{align}  \nonumber
	    \phi_{n} = & \cos \frac{\theta}{2} \phi_{up} + \sin \frac{\theta}{2} \phi_{dn}  \\ \nonumber
	    		= & \cos \frac{\theta}{2} \cos \frac{\varphi}{2} \left[ \frac{1}{\sqrt{2}} \exp (-i \omega_1 \tau), 0, 0,  \frac{1}{\sqrt{2}} \exp (i \omega_1 \tau) \right]^T \\ \nonumber
			 - i & \cos \frac{\theta}{2} \sin \frac{\varphi}{2} \left[ \frac{1}{\sqrt{2}} \exp (-i \omega_1 \tau), 0, 0, - \frac{1}{\sqrt{2}} \exp (i \omega_1 \tau) \right]^T \\ \nonumber
			 + & \sin \frac{\theta}{2} \cos \frac{\varphi}{2} \left[ 0, \frac{1}{\sqrt{2}} \exp (-i \omega_1 \tau), - \frac{1}{\sqrt{2}} \exp (i \omega_1 \tau), 0 \right]^T \\
			 + i & \sin \frac{\theta}{2} \sin \frac{\varphi}{2} \left[ 0, \frac{1}{\sqrt{2}} \exp (-i \omega_1 \tau),  \frac{1}{\sqrt{2}} \exp (i \omega_1 \tau), 0 \right]^T \label{Eq_1_59a}
\end{align}

Consider suppressing the contribution to the superposition for $\phi_n (\tau)$ that comes from the ``negative energy" solution by removing the third and fourth components from each of the four column vectors 
in Eq.~\eqref{Eq_1_59a}, then the common factor $\exp (-i \omega_1 \tau)$ has no effect when calculating dynamic variables and so can be removed. Re-normalizing, this changes $\phi_n$ to the superposition 
\begin{equation}  \nonumber
	\widetilde{\phi}_n =  \cos \frac{\theta}{2}  \exp (-i \frac{\varphi}{2}) [1,0]^T +  \sin \frac{\theta}{2}  \exp (i \frac{\varphi}{2}) [0,1]^T 
\end{equation}
\noindent as in Pauli spin theory. 

It is readily shown from Eq.~\eqref{Eq_1_57a} that $\uvec \cdot \nvec = 0$ so that the electron motion produced by this superposition of spin-up and spin-down states does indeed lie in a plane orthogonal 
to the spin direction $\nvec$. The superposition spin state is therefore physically distinct from each of the four component spin states, which on their own produce motion confined to the $x^1-x^2$ plane. 
Furthermore, from Eq.~\eqref{Eq_1_57a}, $\uvec \cdot \uvec = c^2$ and $\dot{\uvec} \cdot \dot{\uvec} = c^2 \omega_0^2$, so the constraints $C1$ and $C2$ in Eq.~\eqref{C1_C2b} are indeed satisfied. 
Constraint $C3$ is trivially satisfied because here it is just a statement of the rest energy: $\pi_0 u_0 = E = mc^2$, the kinetic energy of the spin motion.
As a final remark, the electron cannot be spinning simultaneously about two distinct spin axes whose directions correspond to $\nvec_1$ and  $\nvec_2$, say, so the Hermitian matrix operators 
$\Sigma_{n_1}$ and  $\Sigma_{n_2}$ corresponding to the spin vectors according to Eq.~\eqref{Eq_1_55a} cannot have common eigenvectors, implying that they are not commutative.

\subsection{Neo-Classical Wave Function and Derivation of Dirac's Wave Equation}

We show that for a \emph{free} electron, the neo-classical Dirac-Schr\"{o}dinger equation~\eqref{Eq_1_23} implies Dirac's wave equation when the Lorentz 
transformation in Eq.~\eqref{Eq_1_10b} is used to give proper time $\tau(x)$ as a function of an observer's 4-vector position $x$ for the electron. 
Define the neo-classical \emph{wave function} $\psi(x) = \phi(\tau(x))$ in terms of the state function  $\phi(\tau)$. 
Consider now the operator $i \hbar \del_{\mu} = i \hbar \frac{\del} {\del x^{\mu}}$ acting on this wave function:
\begin{equation} \label{Eq_1_35}
	i \hbar \del_{\mu} \psi(x) = i \hbar \dot{\phi}(\tau(x)) \del_{\mu} \tau = \del_{\mu} \tau \hat{H} \psi(x) 
\end{equation}
\noindent using Eq.~\eqref{Eq_1_23}. We deduce from the Lorentz transformation  in Eq.~\eqref{Eq_1_6} that $mc^2 \del_{\mu} \tau = \pi_{\mu}$ since 
$d\tau = \del_{\mu} \tau~dx^{\mu}$ and $dx^{\mu}$ $(\mu = 0,1,2,3)$ is arbitrary, so 
\begin{equation} \label{Eq_1_36}
	i \hbar \del_{\mu} \psi(x) =  \frac{\pi_{\mu}} {mc^2} \hat{H} \psi(x)
\end{equation}
\noindent and 
\begin{equation} \label{Eq_1_37}
	\bar{\psi}(x)~(i \hbar \del_{\mu}) \psi(x) = \frac{\pi_{\mu}} {mc^2} \bar{\psi} \hat{H} \psi = \pi_{\mu}
\end{equation}
\noindent using Eq.~\eqref{Eq_1_30}. Thus, the operator $\hat{\pi}_{\mu} = i \hbar \del_{\mu}$ acting on the neo-classical wave function gives the kinetic momentum, as in QM theory. 

From Eq.~\eqref{Eq_1_36}:
\begin{equation} \label{Eq_1_38}
	c \gamma^{\mu} \hat{\pi}_{\mu} \psi(x) = \frac{1} {mc^2} \hat{H}^2 \psi(x) 
\end{equation}
\noindent Applying the identity in Eq.~\eqref{Eq_1_39}, we get the covariant form of \emph{Dirac's wave equation} ({\color{Green}Bjorken and Drell (1964)}): 
\begin{equation} \label{Eq_1_40}
	\widetilde{H} \psi(x) = \hat{u}^{\mu} \hat{\pi}_{\mu} \psi(x) = mc^2 \psi(x) 
\end{equation}
\noindent showing that it corresponds to an eigenvalue equation for the Hamiltonian energy operator $\widetilde{H} =  \hat{u}^{\mu} \hat{\pi}_{\mu}$. 
Dirac's equation can also be interpreted as an operator form of the energy equation $C3$ in Eq.~\eqref{Eq_1_7}. 
It is a linear partial differential equation for the neo-classical wave function $\psi(x)$ whereas the set of ordinary differential equations in Eq.~\eqref{Eq_1_31} involving the state function $\phi(\tau)$ is nonlinear in general. 
However, since the partial differential equation is for the field $\psi(x)$ over the space of the observer's 4-coordinates, this comes at the cost of no longer tracking the space-time path of the electron. 
Therefore, information present in the original equations of motion is lost in Dirac's wave equation, implying that it no longer gives a complete description of the electron. 

From Eq.~\eqref{Eq_1_40}:
\begin{equation} \label{Eq_1_40b}
	\widetilde{H}^2 \psi = \widetilde{H} (mc^2 \psi) = (mc^2)^2 \psi
\end{equation}
\noindent where 
\begin{equation} \label{Eq_1_40c}
	\widetilde{H}^2 =  \hat{u}^{\mu}  \hat{u}^{\nu} \hat{\pi}_{\mu} \hat{\pi}_{\nu} = - (c \hbar)^2 \partial_\mu \partial_\nu (\gamma^\mu \gamma^\nu + \gamma^\nu \gamma^\mu)/2 
				= - (c \hbar)^2 \partial_\mu \partial_\nu g^{\mu\nu} I_4
\end{equation}
\noindent so 
\begin{equation} \label{Eq_1_40d}	
	 \left[ \partial_\mu \partial^\mu + \left( \frac{mc}{\hbar} \right)^2 \right] \psi = 0
\end{equation}
Thus, each component of the neo-classical wave function $\psi$ for the free electron satisfies the Klein-Gordon equation, the QM theory version of Eq.~\eqref{Eq_1_41}.


We can get Dirac's equation for the case of an electron in an em-field by using the usual minimal coupling from QM theory, as did {\color{Green}Dirac (1928)} after developing 
his equation for the free electron. 
Thus, the kinetic momentum operator in Eq.~\eqref{Eq_1_40} becomes $\hat{\pi}_{\mu} = i \hbar \del_{\mu} - qA_{\mu}$, so that now $\hat{p}_{\mu} = i \hbar \del_{\mu}$ is the canonical momentum operator. 
However, minimal coupling in the neo-classical Lagrangian function, which is given in Eq.~\eqref{Eq_1_11}, does not produce the same Dirac equation as this QM minimal 
coupling prescription. 
Indeed, the derivation presented for Dirac's wave equation does not work in the presence of an em-field by setting the operator $i \hbar \del_{\mu}$ equal to $\hat{p}_{\mu}$ 
rather than $\hat{\pi}_{\mu}$. 
Furthermore, the neo-classical Dirac-Schr\"{o}dinger equation~\eqref{Eq_1_23} for the state function needs to be supplemented by Newton's Second Law, as in 
Eq.~\eqref{Eq_1_31}, 
because the Hamiltonian $\hat{H}(\tau)$ in Eq.~\eqref{Eq_1_23} depends on the momentum $\pi(\tau)$, whereas the Dirac Hamiltonian $\widetilde{H}$ does not. 
Newton's Second Law is implicit in Dirac's wave equation under the minimal coupling prescription; it appears, for example, in the Heisenberg equations from Dirac's equation 
in proper time ({\color{Green}Barut (1987)}) and is identical to the last equation in  Eq.~\eqref{Eq_1_34}.

From the Lorentz transformation in Eq.~\eqref{Eq_1_10b}, setting $\omega_1 = \frac{mc^2}{\hbar} = \frac{1}{2} \omega_0$, 
\begin{equation} \label{Eq_1_53}
	 \theta(x) \equiv \omega_1 \tau(x) = (\pi*x) / \hbar = \left( Et - \Pvec \cdot \xvec \right) / \hbar
\end{equation}
\noindent Substituting this in Eq.~\eqref{Eq_1_42}, the solution for the neo-classical wave function for the free electron with global momentum $\pi$ is:
\begin{equation} \label{Eq_1_54}
	\psi(x) = \phi(\tau(x)) = \left[ \cos \theta(x) I_4 - \frac{i}{mc^2} \sin \theta(x) \hat{H} \right] A
\end{equation}
\noindent for arbitrary energy-normalized $A \in {\cal C}^4$. 
This equation shows that the phase $\theta(x)$ defined in Eq.~\eqref{Eq_1_53} gives the wave function $\psi(x)$ an apparent plane-wave characteristic with the well-known 
de Broglie frequency-energy and wavenumber-momentum relations, as well as an apparent wave propagation speed of $c^2/| \Vvec | > c$ 
when viewed by an observer fixed with respect to the reference frame $X_o$. 
Also, $2 \theta(x)$ gives the position of the free electron in its periodic motion about the spin center (to within an unknown additive constant) when it has inertial observer 
coordinates $x = (ct, \xvec)$. Furthermore, $S(x) = \theta(x) \hbar = \pi*x$ is the action function from classical electron mechanics.

Since Dirac's wave equation is linear, a superposition of solutions of the form of Eq.~\eqref{Eq_1_54},  each with a distinct  global momentum $\pi$, is also a solution. 
Each such solution has a distinct proper time $\tau(x)$ according to Eq.~\eqref{Eq_1_53} when the electron has space-time coordinates $x$ relative to $X_o$.
Such a superposition of neo-classical wave functions has no clear interpretation in the present theory whereas in QM theory,  a superposition of wave functions over all possible global momenta 
is postulated to provide a probability distribution (at each observer time $t$) over the electron's spatial coordinates in a region known to contain the electron.

Expressing the sine and cosine terms in Eq.~\eqref{Eq_1_54} as complex exponentials: 
\begin{align}  \nonumber
	\psi(x) & = \phi(\tau(x)) = \left[ \frac{1}{2} \left(I_4 + \frac{1}{mc^2} \hat{H} \right) \exp (-i \theta(x)) + \frac{1}{2} \left(I_4 - \frac{1}{mc^2} \hat{H} \right) \exp (i \theta(x)) \right] A \\
		  & =  \psi^+(x) + \psi^-(x)  \label{Eq_1_55}
\end{align}
\noindent where from the results of the previous subsection, 
\begin{equation} \label{Eq_1_56}
	\psi^+ (x) = \exp (- i\theta(x)) A^+, \hspace{0.2 in} \psi^- (x) = \exp ( i\theta(x)) A^-
\end{equation}
\noindent are the so-called \emph{positive energy} and \emph{negative energy} wave function solutions, although they both have positive energy  
since they are just the previous neo-classical state functions $\phi^+$ and $\phi^-$ in Eq.~\eqref{Eq_1_46a} with $\tau$ transformed using Eq.~\eqref{Eq_1_53}.
Eq.~\eqref{Eq_1_55} agrees with the wave function solution of Dirac's wave equation for the free electron given in textbooks (e.g. {\color{Green}Bjorken and Drell (1964)}). 

Recall that $\hat{H}$ incorporates the specified global kinetic momentum $\pi$ and it is replaced by the operator $\hat{\pi}$ in $\widetilde{H}$. 
For the free electron case considered here, Eq.~\eqref{Eq_1_56} shows that $\hat{\pi}_\mu \psi^+ (x) = \pi_\mu \psi^+ (x)$ whereas $\hat{\pi}_\mu \psi^- (x) = - \pi_\mu \psi^- (x)$, 
so $\widetilde{H} \psi^+(x) = \hat{H} \psi^+(x)$ and $\widetilde{H} \psi^-(x) = - \hat{H} \psi^-(x)$. Dirac's equation therefore gives $\hat{H} \psi^+(x) = mc^2 \psi^+(x)$ and $\hat{H} \psi^-(x) = - mc^2 \psi^-(x)$, 
which is consistent with the result in the previous subsection  that $A^+$ and $A^-$ are eigenvectors of $\hat{H}$ belonging to the two-dimensional eigenspaces corresponding to eigenvalues $mc^2$ and $-mc^2$, respectively. 

When the free electron is at rest globally relative to the observer's inertial reference frame at $O$, then the global 3-vector momentum $\Pvec = \bf{0}$, the kinetic energy is just the spin energy $E=mc^2$ and 
from Eq.~\eqref{Eq_1_53}, proper time $\tau = t$, the observer time. The wave function $\psi_n$ for the electron spinning about direction $\nvec$ is then identical to the state function $\phi_n$ in Eq.~\eqref{Eq_1_59a}.

\section{On Introducing Probability into the Neo-classical Theory}

The neo-classical spin model presented here has a complete set of physical variables describing the electron but that does not mean that their values will always be known with certainty. 
Any variables whose values are uncertain can be described by an appropriate probability distribution using, for example, the Cox-Jaynes interpretation of probability as a logic for 
quantitative plausible reasoning (e.g. {\color{Green}Beck (2018, 2019)}). 
Although a detailed study of what probability distributions should be added to the present neo-classical theory to bring it closer to QM theory is left for future work,  
a few connections are made in this subsection. 

\subsection{Uncertainty in Initial Conditions and Liouville's Equation}

The most general case for the electron spin model requires a joint PDF (probability density function) over 16 physical variables and it is a function of proper time $\tau$. 
Consider, for example, the PDF $p(x, u, y, \pi | \tau)$ for the state-space form of the equations of motion of the electron given in Eq.~\eqref{Eq_1_17}.
Only the joint PDF for the initial conditions $x(0), u(0), y(0), \pi(0)$ at $\tau = 0$ needs to be specified since $p(x, u, y, \pi | \tau)$ satisfies Liouville's equation from stochastic dynamics theory 
(e.g. {\color{Green}Gardiner (1985)}), a PDE to be solved using as the initial condition $p(x, u, y, \pi | 0)$. 
This PDE applies when the evolution of the state is described by deterministic equations, as in the neo-classical model here, but the initial conditions are uncertain. 

Consider now the state-space form of the equations of motion of the electron given in Eq.~\eqref{Eq_1_31}. For the free electron, if $\phi(0)$ and $\pi(0)$ are specified, then $\phi(\tau)$ and $\pi(\tau) = \pi(0)$ are known. 
There remains the possibility, however, that the initial space-time coordinate $x(0)$ of the electron is uncertain. In this case, the first equation in Eq.~\eqref{Eq_1_31} implies that 
$\dot{x}^{\mu}(\tau)  = \bar{\phi}(\tau) \hat{u}^{\mu} \phi(\tau) = u^{\mu}(\tau)$ is known and so $p(x | \tau, \phi(0), \pi(0))$ satisfies Liouville's equation:
\begin{equation} \label{Eq_1_60}
	\del_\tau p(x |\tau, \phi(0), \pi(0)) + \del_\mu \left[ p(x |\tau, \phi(0), \pi(0)) \, u^\mu (\tau(x)) \right] = 0
\end{equation}
\noindent where $\tau(x)$ is the proper time for the electron when it is at the space-time point $x$, which for a free electron with global momentum $\pi$ is given by  $\tau(x) = \tau_0 + (\pi * x)/ (mc^2)$.  
Here $\tau_0$ is an arbitrary constant that would be zero if $\tau=0$ is chosen to occur when $x=0$, that is, at the origin $\xvec=0$ at observer time $t=0$. 
Liouville's equation is to be solved for the joint PDF for $x(\tau)$ for a specified initial PDF $p(x | 0, \phi(0), \pi(0))$ for $x(0)$.

If the uncertainty in $x$ does not change with proper time $\tau$, then the PDF $p(x | \phi(0), \pi(0))$ is a solution of  the ``stationary''  Liouville equation:
\begin{equation} \label{Eq_1_61}
	\del_\mu \left[ p(x | \phi(0), \pi(0)) \, u^\mu (\tau(x)) \right] = 0
\end{equation}
\noindent implying that only an initial distribution $p(x | \phi(0), \pi(0))$ satisfying this PDE gives a stationary PDF.  Define
\begin{equation} \label{Eq_1_62}
	 U^\mu (x) = c \frac{u^\mu (\tau(x))}{u^0 (\tau(x))} = \frac{\dot{x}^\mu (\tau(x))}{\dot{t} (\tau(x))} = \frac{dx^\mu}{dt} \left(\tau(x) \right)
\end{equation}
\noindent implying that $U^0(x) = c$, then the stationary Liouville equation becomes:
\begin{equation} \label{Eq_1_63}
	\del_\mu \left[ p(x | \phi(0), \pi(0)) \, \dot{t} (\tau(x)) \, U^\mu(x) \right] = 0
\end{equation}
\noindent where $U^\mu (x)$ can be expressed using the total velocity operator $\hat{u}^\mu = c \gamma^\mu$ and Dirac's wave function $\psi(x) = \phi(\tau(x))$ for the free electron:
\begin{equation} \label{Eq_1_64}
	 U^\mu (x) =  \frac {\bar{\psi}(x) \hat{u}^{\mu} \psi(x)}{\psi^*(x) \psi(x)}
\end{equation}

Consider now the continuity equation derived from Dirac's wave equation:
\begin{equation} \label{Eq_1_65}
	0 = \del_\mu \left[ j^\mu (x) \right] = \del_\mu \left[ \psi^*(x) \psi(x) \, U^\mu (x) \right] 
\end{equation}
\noindent where the probability current is defined by:
\begin{equation} \label{Eq_1_66}
	 j^\mu (x) =  \bar{\psi}(x) \hat{u}^{\mu} \psi(x) = \psi^*(x) \psi(x) \, U^\mu (x)
\end{equation}
\noindent In QM, $ \psi^*(x) \psi(x)$ is identified with $p(x | \psi)$, viewed as the PDF $p(\xvec, t | \psi)$ for the spatial position $\xvec$ of the electron at time $t$, 
conditional on knowing the wave function $\psi$, so the continuity equation becomes with this identification:
\begin{equation} \label{Eq_1_67}
	\del_\mu \left[ p(x | \psi) \, U^\mu (x) \right] = 0
\end{equation}
\noindent Notice that if $\dot{t} (\tau(x)) = 1$ for all $x$, as in the non-relativistic case, the continuity equation and the stationary Liouville equation are identical with the common solution 
$p(x | \phi(0), \pi(0)) = p(x | \psi)$ that is stationary with respect to $\tau = t$ and so is just $p(\xvec | \psi)$. 
In general, however, $\dot{t} (\tau(x))$ is highly oscillatory and $p(x | \phi(0), \pi(0)) \, \dot{t} (\tau(x))$ cannot be identified with the PDF $p(\xvec, t | \psi)$ to give agreement between 
Liouville's equation and the continuity equation with the stated QM identification of $ \psi^*(x) \psi(x)$ as a PDF. 

For the free electron, Liouville's equation in Eq.~\eqref{Eq_1_61} has the solution $p(x)$ constant over a finite space-time region ${\cal V}$ where it can be normalized so that 
$\int_{\cal V} p(x) dx = 1$. This is evident from the fact that the continuity equation derived from Dirac's wave equation for $\psi(x) = \phi(\tau(x))$ is:
\begin{equation} \label{Eq_1_68}
	\del_\mu \left[ u^\mu (\tau(x)) \right] = \del_\mu \left[ \bar{\psi}(x)  \, \hat{u}^\mu \psi(x)  \right] = 0
\end{equation}
\noindent This equation also follows from Section 4.4:
\begin{equation} \label{Eq_1_69}
	\del_\mu \left[ u^\mu (\tau(x)) \right] = \dot{u}^\mu (\tau(x)) \del_\mu \tau(x) = \frac{1}{mc^2} \dot{u}^\mu \pi_\mu = 0
\end{equation}
\noindent using Eqs.~\eqref{Eq_1_1} and~\eqref{Eq_1_8}, or Eq.~\eqref{Eq_1_18} and the antisymmetry of the spin tensor.
Thus, for any proper time $\tau$, the uncertainty in the electron's space-time coordinates corresponds to a uniform distribution over ${\cal V}$; 
in particular, the PDF for the free electron's spatial coordinates, $p(\xvec)$, is constant, as in QM theory.

\subsection{Uncertainty in Spin Measurement Outcomes and Malus's Law}

As noted at the end of the previous section, the  wave function $\psi_n(x) $ for the free electron spinning about direction $\nvec$ is identical to the state function 
$\phi_n(\tau)$ given by the superposition in Eq.~\eqref{Eq_1_59a}. Thus, applying Born's rule from QM, the coefficients $\cos \frac{\theta}{2}$ and $\sin \frac{\theta}{2}$ 
when squared give the probabilities $P_{up}$ and $P_{dn}$ of getting spin-up and spin-down states, respectively, upon exit from a Stern-Gerlach device, 
given that the electron is in a state of spin about the direction $\nvec$ upon entry:
\begin{equation} \label{Eq_1_57}
	P_{up} = \cos^2 \frac{\theta}{2} = \frac{1}{2} \left( 1 + \nvec \cdot {\bf e}_3 \right),  \hspace{0.1in} P_{dn} = \sin^2 \frac{\theta}{2} = \frac{1}{2} \left( 1 - \nvec \cdot {\bf e}_3 \right)
\end{equation}
\noindent This is Malus's Law for spin-1/2 particles (e.g. {\color{Green}W\'odikiewicz (1985)}).

It is interesting to note that the expressions for $u^i (i=1,2)$ in Eq.~\eqref{Eq_1_57a} for the electron spinning about direction $\nvec$ upon entering the Stern-Gerlach device 
are equal to the predicted means of these velocities for the electron emerging from the device, since the expressions can be interpreted as the sum $P_{up} u^i_{up} + P_{dn} u^i_{dn}$ 
where $u^i_{up}$ is given by Eq.~\eqref{Eq_1_50a} and $u^i_{dn}$ is given by Eq.~\eqref{Eq_1_51a}. However, $u^3$ in Eq.~\eqref{Eq_1_57a} does not agree with 
the mean velocity in the ${\bf e}_3$ direction, which is obviously zero because $u^3_{up}$ and $u^3_{dn}$ are both zero.

Born's rule could be incorporated into the neo-classical spin theory to extend it to spin ``measurement" outcomes. 
This gives the state function a role beyond being just a mathematical device for conveniently solving the equations in Eq.~\eqref{Eq_1_18}, or their equivalent in Eq.~\eqref{Eq_1_1}, 
suggesting that it captures some additional physics. The term ``measurement" of spin here, although commonly used, is misleading because it suggests that a 
pre-existing spin direction is being measured; however, as inferred from Malus's Law, the incoming spin direction is rotated by the magnetic field of the Stern-Gerlach device 
into either the up or the down direction, albeit in a deterministically unpredictable way. 
This inference is consistent with the picture that emerges from the analyzed motion of a charged particle with spin as it passes through a Stern-Gerlach device 
where the analysis is based on Bohmian mechanics and Pauli's non-relativistic wave equation ({\color{Green} Dewdney, Holland and Kyprianidas (1986)}). 
Their results show that if the spin direction relative to the axes of the device is known upon entry, the uncertainty in whether the spin direction upon exit is up or down 
is due to the uncertainty in the exact lateral location of the electron when it enters. 

There has been a long interest in examining the possibility that this indeterminacy in spin outcomes is due to ``hidden variables" (e.g. {\color{Green}Bell (1987), 
Brunner et al. (2014)}). In the context of the neo-classical model presented here, since the spin direction is an input in Malus's Law, this leaves the spin phase, 
which specifies where the electron is in its spin cycle, as a possible hidden variable that could influence the outcome. 
An analogous setting that comes to mind is the flipping of a coin. If the initial conditions for the center of mass and angular velocity of the coin where known to a very high 
precision, then the outcome of a head or a tail could be predicted deterministically; however, the  angular velocity is so high that the outcome is very sensitive to any 
uncertainty in the values of the initial conditions, so that in practice, it is uncertain. 
The spin frequency $\omega_0$ of the electron is ultra-high but in contrast to the flipping of the coin, it seems unlikely that the spin outcome for the electron emerging 
from the Stern-Gerlach device could be dependent on the spin phase. 
However, since the magnetic force on the electron depends on its location in the inhomogeneous magnetic field of the device, a dependence on the initial 
spatial coordinates of the electron as it enters the device is to be expected, as in the Bohmian analysis. It is doubtful, though, that the neo-classical spin theory 
can explain why the initial spin direction is always rotated to an up or down direction as it passes through the Stern-Gerlach device. The Bohmian analysis of 
{\color{Green} Dewdney, Holland and Kyprianidas (1986)} reveals a mysterious "quantum torque" that effects the spin direction, even in the absence of a magnetic field.

\section{Concluding Remarks}

The neo-classical theory presented here gives a classical relativistic mechanics model for an electron that explains its spin motion 
as an inherent property of its space-time trajectory as a point particle. 
The fourth-order equation of motion in proper time is obtained by simply adding a spin energy term to the covariant Lagrangian function of special relativity. 
This space-time equation of motion is equivalent to a second-order equation for a local motion of the electron about a point, its spin center, 
coupled to an equation for the motion of this center that corresponds to Newton's Second Law in proper time for the electron's global motion. 

The theory provides an underlying reality for many mysterious features of spin that is missing in QM (quantum mechanics), including: 

\begin{itemize} 
	\item[--] The physical nature of the spin of the electron, which is known to not be a simple rotation of it. The theory gives it as 
	a perpetual local motion about a globally moving spin center that is inherent in the electron's space-time trajectory as a ``point'' particle.
	For a free electron, this spin motion is circular about a constant spin axis at the ultra-high zitterbewegung frequency when it is viewed from a spin-center reference frame.
	\item[--] The phenomenon of zitterbewegung revealed by Schr\"{o}dinger's analysis of Dirac's wave equation for the free electron. It is a manifestation of 
	its local spin motion, which is also the mechanism for de Broglie's ``internal clock'' of the electron. 
	\item[--] An apparent plane-wave characteristic of the electron's motion, consistent with de Broglie's wave theory. It is just the local spin motion viewed 
	through the Lorentz transformation by an observer fixed with respect to another reference frame. 
	\item[--] The dependence of the electron's kinetic energy on an observer's reference frame is accounted for by its storage as the spin rotational energy (like a flywheel) 
	with an observed angular speed that depends on the clock rate for that reference frame.
	\item[--] The eigenvalues of Dirac's velocity operators have magnitude $c$ (the speed of light) because the total speed of the electron is always $c$. The global speed  
	of the spin center is always sub-luminal with the difference between the speeds of the electron and spin center being accounted for by the local spin motion.
	\item[--] The physical mechanism behind the generation of electric and magnetic dipole energies. It is due to the action of the em-force on the spin motion.
	\item[--] The physical nature of the QM spin tensor. It represents the angular momentum from the electron's total velocity about its spin center.
	\item[--] An interpretation of the wave function satisfying Dirac's relativistic wave equation for the free electron. It is given by the neo-classical state function from the Barut-Zanghi 
	``classical Dirac'' theory for an electron after using the Lorentz transformation to express proper time in the state function in terms of an inertial observer's space-time coordinates 
	for the electron. This straightforward connection between the state and wave functions does not seem to hold, however, when the electron is interacting with an electro-magnetic field. 
	\item[--] The nature of the superposition of the spin-up and spin-down wave functions that reveals the spin motion of a free electron about a specified spin axis from Dirac's velocity and spin operators. 
\end{itemize}

Based on the presented theory, one can make the following interpretation of the Dirac wave function $\psi_n(x) = \phi_n(\tau(x))$ in Eq.~\eqref{Eq_1_59a}. 
As shown in Section 4.3, $\psi_n(x)$ gives deterministically the dynamics of the free electron with spin direction $\nvec$ where Dirac's spin operator gives an angular 
momentum for the electron of $\frac{1}{2} \hbar$ about $\nvec$ and his velocity operator shows that this comes from circular motion of the electron at speed $c$ and an 
angular frequency $\omega_0$ in a plane perpendicular to $\nvec$. Thus, the wave function has an \emph{ontic} role. 
If the electron subsequently enters a Stern-Gerlach device and interacts with its magnetic field, then which of the two output states, spin up or spin down, will occur is 
uncertain and Born's rule gives the probability for each outcome as in Eq.~\eqref{Eq_1_57}. Thus, the wave function then has an \emph{epistemic} role. 

Although the theory exhibits magnetic and electric dipole energies that are a half of Dirac's theory, it does so by deriving an energy equation with a spin-field interaction term that comes 
from the em-force acting on the spin component of the electron's motion, rather than invoking the usual minimal coupling prescription assumed in QM, as Dirac did to extend 
his free-electron theory. The electric dipole energy term explains the spin-orbit coupling for an electron moving in an electric field such as that due to the proton in the hydrogen atom. 

As far as an electron's free motion is concerned, the theory may be viewed as a classical mechanics alternative to QM that explains its motion at a deeper level in terms of dynamic variables.
Its dynamics also have some close connections to QM theory. For example, the neo-classical equations of motion written in spin tensor form as in Eq.~\eqref{Eq_1_18} 
are identical to those in the Heisenberg picture of the proper time Dirac equation except that actual dynamic variables are used where 
the corresponding operators appear (see Eq.~\eqref{Eq_1_18} and Eq.~\eqref{Eq_1_34}. 
Also, Dirac's equation for the free electron can be derived from the operator form of the neo-classical theory. However, the theory does not provide, 
in its current form, an explanation of the stationary energy states found in the Schr\"{o}dinger picture for Dirac's equation. 
It remains to be seen whether the theory can be further developed to better explain the electron's interactions with em-fields. 
It seems likely that this extension will need to go hand in hand with the way probability is introduced in order to describe uncertainty in the values 
of the dynamic variables so that the theory is consistent with the probabilistic predictions of QM theory.



\appendix


\section{}

Here the equations of motion in Eq.~\eqref{Eq_1_18} are shown to imply the pair of equations in Eq.~\eqref{Eq_1_1} and so their state-space re-formulation in Eq.~\eqref{Eq_1_17}. 
Differentiating the equation for the total acceleration in Eq.~\eqref{Eq_1_18}:
\begin{equation} \label{Eq_A_1}
	\ddot{u}^\mu = \frac{4c^2}{\hbar^2} \left[ \dot{S}^{\mu\nu} \pi_\nu + S^{\mu\nu} \dot{\pi}_\nu \right]
\end{equation}
\noindent Using Eq.~\eqref{Eq_1_18} to substitute for the two derivatives:
\begin{equation} \label{Eq_A_2}
	\ddot{u}^\mu = \frac{4c^2}{\hbar^2} \left[ \pi^\mu (u^\nu \pi_\nu) - u^\mu (\pi^\nu \pi_\nu) + q S^{\mu\nu} F_{\nu\rho} u^\rho \right]
\end{equation}
\noindent From the definition of the spin tensor in Eq.~\eqref{Eq_1_13}:
\begin{equation} \label{Eq_A_3}
	q S^{\mu\nu} F_{\nu\rho} u^\rho  = qm (z^\nu F_{\nu\rho} u^\rho) u^\mu - qm (u^\nu F_{\nu\rho} u^\rho) z^\mu
\end{equation}
\noindent The last term on the right is zero because of the anti-symmetry of the em-field tensor. The first term on the right is $m \Phi u^\mu$ from Eq.~\eqref{Eq_1_21}. Substituting into Eq.~\eqref{Eq_A_2}:
\begin{equation} \label{Eq_A_4}
	\ddot{u}^\mu = \frac{4mc^2}{\hbar^2} (u^\nu \pi_\nu) \dot{y}^\mu - \frac{4mc^2}{\hbar^2} \left[ \frac{1}{m} \pi^\nu \pi_\nu - \Phi  \right] u^\mu 
\end{equation}

Define $G(\tau) = z * \pi$, then using $C3$ in Eq.~\eqref{Eq_1_7} and the definition of $\Phi$ after Eq.~\eqref{Eq_1_9}:
\begin{equation} \label{Eq_A_5}
	\dot{G} = \pi * \dot{z} + \dot{\pi} * z = \pi * u - \pi *  \dot{y} +  f * z = mc^2 - \frac{1}{m} \pi * \pi + \Phi   
\end{equation}
\noindent Substituting Eq.~\eqref{Eq_A_5} into Eq.~\eqref{Eq_A_4} while using $C3$ and the definition of $\omega_0$:
\begin{equation} \label{Eq_A_6}
	\ddot{u}^\mu = - \omega_0^2 \dot{z}^\mu + \frac{2 \omega_0}{\hbar} \dot{G} u^\mu 
\end{equation}

An alternative expression for $\ddot{u}^\mu$ can be obtained by substituting the definition of the spin tensor in  Eq.~\eqref{Eq_1_13} into the equation for the total acceleration in Eq.~\eqref{Eq_1_18} and using $C3$:
\begin{equation} \label{Eq_A_7}
	\dot{u}^\mu = \frac{4c^2}{\hbar^2} S^{\mu\nu} \pi_\nu =  - \omega_0^2 z^\mu + \frac{2 \omega_0}{\hbar} G u^\mu 
\end{equation}
\noindent and then differentiating:
\begin{equation} \label{Eq_A_8}
	\ddot{u}^\mu = - \omega_0^2 \dot{z}^\mu + \frac{2 \omega_0}{\hbar} \dot{G} u^\mu +  \frac{2 \omega_0}{\hbar} G  \dot{u}^\mu 
\end{equation}
\noindent For consistency between these two equations for $\ddot{u}^\mu$, we must have $G  \dot{u}^\mu = 0$ for each $\mu$, implying $G=0$, and so from Eq.~\eqref{Eq_A_6} or Eq.~\eqref{Eq_A_7}:
\begin{equation} \label{Eq_A_9}
	\dot{u}^\mu = - \omega_0^2 z^\mu
\end{equation}
\noindent This proves that under the constraint $C3$, Eq.~\eqref{Eq_1_18} imply both $G=0$ and the pair of equations in Eq.~\eqref{Eq_1_1} (Newton's Second Law is common to both sets of equations). 
Substituting $G=0$ into Eq.~\eqref{Eq_A_5} gives the energy equation in Eq.~\eqref{Eq_1_9}, which was proved in Section 2 by deriving $z * \pi=G=0$ in Eq.~\eqref{Eq_1_8} 
from Eq.~\eqref{Eq_1_1}. 
In Section 3, it was shown that the equations in Eq.~\eqref{Eq_1_1} imply Eq.~\eqref{Eq_1_18}. Thus, the two sets of equations are equivalent. 

It remains to show that the respective initial conditions for the two equivalent sets of equations imply each other. For the equations in Eq.~\eqref{Eq_1_1}, and their state-space form Eq.~\eqref{Eq_1_17}, 
the initial state $(x(0), u(0), y(0), \pi(0))$ must be specified. To find the corresponding initial conditions for Eq.~\eqref{Eq_1_18}, we need only find $S(0)$. But $z(0)=x(0)-y(0)$ is known, as is $u(0)=\dot{x}(0)$, 
so from the definition of the spin tensor in Eq.~\eqref{Eq_1_13}, $S(0)$ is known. Conversely, if the initial conditions $(x(0), u(0), S(0), \pi(0))$ for Eq.~\eqref{Eq_1_18} are specified, then we 
need only find $y(0)$. But from Eq.~\eqref{Eq_1_19}, $z(0)$ is known and so $y(0)=x(0)-z(0)$ is too.

\section{}

Here we derive the expression for the total velocity from its operator and the free electron solution $\phi(\tau)$ of the neo-classical Dirac-Schr\"{o}dinger equation for the state function.
This expression was presented without derivation in {\color{Green} Barut and Zanghi (1984)} in their Eq. (4). 

From Eq.~\eqref{Eq_1_42},
\begin{equation}  \label{Eq_B_1}
	\phi(\tau) = \left[ \cos (\omega_1 \tau) I_4 - \frac{i}{mc^2} \sin (\omega_1 \tau) \hat{H} \right] \phi(0)
\end{equation}
\noindent  Therefore, 
\begin{align}  \nonumber
	\bar{\phi}(\tau)  & = \bar{\phi}(0) \left[ \cos (\omega_1 \tau) \gamma^0 + \frac{i}{mc^2} \sin (\omega_1 \tau)  \gamma^0 \hat{H}^* \right]  \gamma^0 \\ 
				& = \bar{\phi}(0) \left[ \cos (\omega_1 \tau) I_4 + \frac{i}{mc^2} \sin (\omega_1 \tau) \hat{H} \right] \label{Eq_B_2}
\end{align}
\noindent since $(\gamma^0)^2 = I_4$ and $\gamma^0 \hat{H}^* \gamma^0 = \hat{H}$.

\noindent Using the velocity operator,
\begin{align}  \nonumber
	& u^\mu(\tau) = \bar{\phi}(\tau) c \gamma^\mu \phi(\tau) \\
	& = \bar{\phi}(0) \left[ c \cos^2 (\omega_1 \tau) \gamma^\mu + \frac{i}{2mc} \sin (2 \omega_1 \tau) \left[ \hat{H}, \gamma^\mu \right] 
		 +  \frac{c}{(mc^2)^2} \sin^2 (\omega_1 \tau) \hat{H}  \gamma^\mu \hat{H} \right] \phi(0)  \label{Eq_B_3}
\end{align}
\noindent From Eq.~\eqref{Eq_1_32}, the commutator $\left[ \hat{H}, \gamma^\mu \right] =  - \frac{i\hbar}{c} \hat{\dot{u}}^{\mu}$.
Furthermore, using the identity for the Dirac matrices: 
\begin{align}  \nonumber
	\hat{H}  \gamma^\mu \hat{H} & =  c^2 \pi_\nu \pi_\sigma \gamma^\nu \gamma^\mu \gamma^\sigma  \\ \nonumber
	& = - c^2 \pi_\nu \pi_\sigma \gamma^\nu \gamma^\sigma \gamma^\mu + 2 c^2 \pi_\nu \pi_\sigma g^{\mu\sigma} \gamma^\nu  \\ \nonumber
	& = - \frac{c^2}{2} \pi_\nu \pi_\sigma \left( \gamma^\nu \gamma^\sigma +  \gamma^\sigma \gamma^\nu \right)  \gamma^\mu + 2c \pi^\mu \hat{H}  \\ 
	& = - c^2 \pi_\nu \pi^\nu  \gamma^\mu + 2c \pi^\mu \hat{H}  =  - \left( mc^2 \right)^2  \gamma^\mu + 2c \pi^\mu  \hat{H}  \label{Eq_B_4}
\end{align}
\noindent Substituting these results into Eq.~\eqref{Eq_B_3}:
\begin{align}  \nonumber
	u^\mu(\tau) & = \bar{\phi}(0) \left[ \left( c \gamma^\mu- \frac{1}{(mc)^2} \pi^\mu \hat{H} \right) \cos(\omega_0 \tau) + \frac{1}{\omega_0} \hat{\dot{u}}^{\mu} \sin (\omega_0 \tau) + 
	\frac{1}{(mc)^2} \pi^\mu \hat{H} \right] \phi(0)  \\
	& = \left( u^\mu(0) - \frac{1}{m} \pi^\mu \right) \cos(\omega_0 \tau) + \frac{1}{\omega_0} \dot{u}^{\mu}(0) \sin (\omega_0 \tau) + \frac{1}{m} \pi^\mu  \label{Eq_B_5}
\end{align}
\noindent using $\bar{\phi}(0) \hat{H} \phi(0) = mc^2$ from Eq.~\eqref{Eq_1_30}. 
This equation is identical to the solution for the total velocity given in the first equation of Eq.~\eqref{Eq_1_4}.

\section{}

Here the total velocity components $u^j$ and spin components $s^j$ $(j = 1,2,3)$ are given for the state function $\phi(\tau)$ in Eq.~\eqref{Eq_1_48a}. 
From Eq.~\eqref{Eq_1_24}, the 4-vector velocity components are given by $u^{\mu} = \bar{\phi} \hat{u}^{\mu} \phi = c \bar{\phi} \gamma^{\mu} \phi$. 
Substituting for the Dirac matrices and for $\phi(\tau)$ from Eq.~\eqref{Eq_1_48a}:
\begin{align}  \nonumber
	u^0 & = c \phi^* \phi = c (|A_1|^2 + |A_2|^2 + |A_3|^2 + |A_4|^2)  \\  \nonumber
	u^1 & = c ( \bar{A}_1A_4 + \bar{A}_2A_3 ) e^{i \omega_0 \tau}  + c ( A_1\bar{A}_4 + A_2\bar{A}_3 ) e^{- i \omega_0 \tau}  \\  \nonumber
	u^2 & = - i c ( \bar{A}_1A_4 - \bar{A}_2A_3 ) e^{i \omega_0 \tau}  + i c ( A_1\bar{A}_4 - A_2\bar{A}_3 ) e^{- i \omega_0 \tau} \\ 
	u^3 & = c ( \bar{A}_1A_3 - \bar{A}_2A_4 ) e^{i \omega_0 \tau}  + c ( A_1\bar{A}_3 - A_2\bar{A}_4 ) e^{- i \omega_0 \tau}  \label{Eq_C_1}
\end{align}
\noindent  where an overbar is used to denote a complex conjugate of a scalar.

From Eq.~\eqref{Eq_1_14}, the spin vector $\svec = (S^{32}, S^{13}, S^{21})$ and so the corresponding operator is $\hat{\svec} = (\hat{S}^{32}, \hat{S}^{13}, \hat{S}^{21})$ 
where $\svec = \bar{\phi} \hat{\svec} \phi$. Since $\hat{S}^{\mu\nu} = i \frac{\hbar}{2} \gamma^\nu \gamma^\mu$ for $\mu \neq \nu$, 
\begin{equation} \label{Eq_C_2}
	\hat{s}^1 = i \frac{\hbar}{2} \gamma^2 \gamma^3 = - i \frac{\hbar}{2} 
		\left[  \begin{array}{cc} 
		\sigma^2 \sigma^3 & 0 \\
		0 & \sigma^2 \sigma^3  
\end{array} \right]
\end{equation}
\noindent where the Pauli spin matrices satisfy $\sigma^2 \sigma^3 = i \sigma^1$ so:
\begin{equation} \label{Eq_C_3}
	\hat{s}^1 = \frac{\hbar}{2} 
		\left[  \begin{array}{cc} 
		\sigma^1 & 0 \\
		0 & \sigma^1  
	\end{array} \right]
\end{equation}
Using a similar approach for $\hat{s}^2$ and $\hat{s}^3$, we have for $j =1,2,3$:
\begin{equation} \label{Eq_C_4}
	s^j = \bar{\phi} \hat{s}^j \phi = \frac{\hbar}{2} \phi^* \Sigma^j \phi 
\end{equation}
\begin{equation} \label{Eq_C_5}
	\Sigma^j = \left[  \begin{array}{cc} 
			\sigma^j & 0 \\
			0 & - \sigma^j  
	\end{array} \right]
\end{equation}
Substituting for the Pauli matrices and for $\phi(\tau)$ from Eq.~\eqref{Eq_1_48a}:
\begin{align}  \nonumber
	s^1 & = \frac{\hbar}{2} ( \bar{A}_1A_2 + A_1\bar{A}_2 - \bar{A}_3A_4 - A_3\bar{A}_4 )  \\  \nonumber
	s^2 & = - i \frac{\hbar}{2} ( \bar{A}_1A_2 - A_1\bar{A}_2 - \bar{A}_3A_4 + A_3\bar{A}_4 )  \\ 
	s^3 & = \frac{\hbar}{2} ( |A_1|^2 - |A_2|^2 - |A_3|^2 + |A_4|^2)  \label{Eq_C_6}
\end{align}
\noindent  These spin components are constant with respect to proper time $\tau$, as expected for a free electron.




\section*{References}
\bibliographystyle{elsarticle-harv.bst}
\bibliography{xxx}

\noindent A.O. Barut (1987), Electron as a radiating and spinning dynamical system and discrete internal quantum systems, Phys. Scripta, 35, 229-232. \\

\noindent A.O. Barut and A.J. Bracken (1982), Exact solutions of the Heisenberg equations and zitterbewegung of the electron in a constant uniform magnetic field, Austr. J. Phys. 35, 353-370. \\

\noindent A.O. Barut and W.D. Thacker (1985), Zitterbewegung of the electron in external fields, Phys. Rev. D 31, 2076-2088. \\

\noindent A.O. Barut and N. Zanghi (1984), Classical model of the Dirac electron, Phys. Rev. Lett. 52, 2009-2012. \\

\noindent G. Baym (1981), Lectures in Quantum Mechanics, 9th printing, W.A. Benjamin Inc., Reading, MA, USA. \\

\noindent J.L. Beck (2018), Contrasting implications of the frequentist and Bayesian interpretations of probability when applied to quantum mechanics theory, arXiv:1804.02106. \\

\noindent J.L. Beck (2019), Invalidity of the standard locality condition for hidden-variable models in Bohm-EPR experiments, Int. J. Quantum Foundations, 5, 115-133. \\

\noindent J.S. Bell (1987), Speakable and Unspeakable in Quantum Mechanics, Cambridge University Press, Cambridge, UK. \\

\noindent J.D. Bjorken and S.D. Drell (1964), Relativistic Quantum Mechanics, McGraw-Hill, New York, USA. \\

\noindent N. Brunner, D. Cavalcanti, S. Pironio, V. Scarani and S. Wehner (2014), Bell nonlocality, Rev. Mod. Phys. 86, 419-478. \\

\noindent O. Consa (2018), Helical solenoidal model of the electron, Progr. Phys. 14, 80-89. \\

\noindent H.C. Corben and P. Stehle (1960), Classical Mechanics, John Wiley and Sons, New York, USA. \\

\noindent A.A. Deriglazov and D.M. Tereza (2019), Covariant version of the Pauli Hamiltonian, spin-induced noncommutativity, Thomas precession, and the precession of spin, Phys. Rev. D 100, 105009. \\

\noindent  C. Dewdney, P.R. Holland and A. Kyprianidas (1986), What happens in a spin measurement?, Phys. Letters A, 119, 259-267. \\

\noindent P.A.M. Dirac (1928), The quantum theory of the electron, Proc. Roy. Soc. London A117, 610. \\

\noindent P.A.M. Dirac (1928), The quantum theory of the electron, Part II, Proc. Roy. Soc. London A118, 351. \\

\noindent P.A.M. Dirac (1958), The Principles of Quantum Mechanics, 4th ed., Oxford Univ. Press, Oxford, UK. \\

\noindent J. Frenkel (1926), Letter to the Editor, Nature 117, 653-654. \\

\noindent C.W. Gardiner (1985), Handbook of Stochastic Methods for Physics, Chemistry and the Natural Sciences, Springer, Berlin, Germany. \\

\noindent H. Goldstein (1959), Classical Mechanics, Addison-Wesley Publishing, Reading, MA, USA. \\

\noindent M. Gouan\`ere, M. Spighel, N. Cue, M.J. Gaillard, R. Genre, R. Kirsch, J.-C. Poizat, J. Remillieux, P. Catillon and L. Roussel (2005), Annales Fond. Louis de Broglie 30, 109. \\

\noindent Z. Grossmann and A. Peres (1963), Classical theory of the Dirac electron, Phys. Rev. 132, 2346-2349. \\

\noindent D. Hestenes (1985), Quantum mechanics from self-interaction, Foundation Phys. 15, 63-87. \\

\noindent D. Hestenes (1990), The Zitterbewegung interpretation of quantum mechanics, Foundation Phys. 20, 1213-1232. \\

\noindent D. Hestenes (1993), Zitterbewegung modeling, Foundation Phys. 23, 365-387. \\


\noindent D. Hestenes (2010), Zitterbewegung in quantum mechanics, Foundation Phys. 40, 1-54. \\

\noindent A.G. Krylovetskii (1978), Canonical formalism and relativistic invariance of the Grossmann-Peres classical electron model, Russian Phys. J. 20, 1322-1326.  \\

\noindent A. Proca (1954), Mecanique du point, J. Phys. Radium 15, 65-72. \\

\noindent F. Riewe (1972), Relativistic classical spinning-particle mechanics, Il Nuovo Cimento 8B, 271-277. \\

\noindent M. Rivas (1989), Classical relativistic spinning particles, J. Math. Phys. 30, 318. \\

\noindent M. Rivas (1994), Quantization of generalized spinning particles. New derivation of Dirac's equation, J. Math. Phys. 35, 3380. \\

\noindent M. Rivas (2001), Kinematical Theory of Spinning Particles, Kluwer, Dordrecht, The Netherlands. \\

\noindent M. Rivas (2003), The dynamical equation of the spinning electron, J. Phys. A: Math. and General 36, 4703. \\

\noindent M. Rivas (2008), The atomic hypothesis: Physical consequences, J. Phys. A: Math. and Theor. 41, 304022. \\

\noindent G. Salesi (2002), Non-Newtonian mechanics, Int. J. Mod. Phys. A 17, 347-374. \\

\noindent G. Salesi (2005), Non-relativistic classical mechanics for spinning particles, Int. J. Mod. Phys. A 20, 2027-2036. \\


\noindent G. Salesi and E. Recami (2000), Effects of spin on the cyclotron frequency for a Dirac electron, Phys. Lett. A 267, 219-224. \\

\noindent E. Schr\"{o}dinger (1930), \"{U}ber die Kr\"{a}fetfreie Bewegung in der relativistischen Quantenmechanik, Sitzungsber. Preuss. Akad. Wiss. Physik-math. K1, 24, 418-428. \\


\noindent G. Spavieri and M. Mansuripur (2015), Origin of the spin-orbit interaction, Physica Scripta 90, 085501. \\

\noindent L.H. Thomas (1926), The motion of the spinning electron, Nature 117, 514. \\

\noindent L.H. Thomas (1927), The kinematics of an electron with an axis, Phil. Mag. 3, 1-22. \\

\noindent G.E. Uhlenbeck and S.A Goudsmit (1925), Naturwissenschaften, 13, 953-954 \\

\noindent G.E. Uhlenbeck and S.A Goudsmit (1926), Spinning electrons and the structure of spectra, Nature, 117, 264-265. \\

\noindent J.W. van Holten (1992), Relativistic time dilation in an external field, Physica A 182, 279-292. \\

\noindent J. Weyssenhoff (1947), Further contributions to the dynamics of spin-particles moving with the velocity of light, Acta Physica Polonica IX, 34-45. \\

\noindent J. Weyssenhoff and A. Raabe (1947), Relativistic dynamics of spin-particles moving with the velocity of light, Acta Physica Polonica IX, 19-25. \\

\noindent K. W\'odikiewicz (1985), Quantum {M}alus' law, Phys Lett A 112, 276--278. \\

\end{document}